# Machine learning-assisted design and explainable optimization of $CdSnP_2$-based integrated solar–photodetector devices

Md. Alamin Hossain Pappu[1,2], Kazi Abrar Shafin[1], Mainul Hossain[3], and Jaker Hossain[1*]

[1]*Photonic & advanced materials Laboratory, Department of Electrical and Electronic Engineering, University of Rajshahi, Rajshahi 6205, Bangladesh.*

[2]*Department of Computer Science & Engineering, Varendra University, Rajshahi 6204, Bangladesh.*

[3]*Department of Electrical and Electronic Engineering, University of Dhaka, Dhaka 1000, Bangladesh.*

## Abstract

$CdSnP_2$-based integrated solar cell–photodetector (SC–PD) devices employing CdS and $CuGaSe_2$ (CGS) as the window and back surface field (BSF) layers, respectively are investigated using a hybrid machine learning (ML)-assisted SCAPS-1D framework. Device optimization is performed by varying the thickness, doping concentration, and defect density of individual layers. SCAPS-generated data are used to train six ML models and one deep learning model, with ensemble-based algorithms exhibiting the highest predictive accuracy. The ML-guided optimization identifies the n-CdS/p-$CdSnP_2$ (CTP)/$p^+$-CGS architecture as the optimum configuration among thirteen candidate structures. Incorporation of a 200 nm CGS BSF layer significantly enhances both photovoltaic and photodetection performance, increasing the efficiency from 20.67% to 32.69%, responsivity from 0.53 $AW^{-1}$ to 0.72 $AW^{-1}$, and detectivity from $2.51\times10^{14}$ Jones to $1.78\times10^{16}$ Jones. SHapley Additive exPlanations (SHAP) analysis reveales that band-offset engineering, particularly at the window/absorber and absorber/BSF interfaces, together with absorber properties, governs device performance. These findings demonstrate the potential of $CdSnP_2$ and the proposed data-driven SCAPS–ML framework for the accelerated design of high-efficiency multifunctional optoelectronic devices.

**Keywords:** Solar Cell-Photodetector (SC-PD), $CdSnP_2$ (CTP), $CuGaSe_2$ (CGS), Responsivity (R), Detectivity ($D^*$), SCAPS-ML framework.

## 1. Introduction

Photonic gadgets play an important role in different technological applications, including solar energy generation, optical communication, information transmission, signal processing, sensing, imaging, and demonstration technologies. Semiconductor-based gadgets such as laser diodes (LDs), light-emitting diodes (LEDs), solar cells, and photodetectors have been established to develop efficiency in energy transformation, and optical data transmission. These gadgets can be broadly categorized into three classes based on their functionality and working principles. Photovoltaic (PV) devices, such as PV cells, alter sunlight into electrical power by generating electron–hole pairs through the photovoltaic influence. On the contrary, Photosensors detect sunlight and convert them into signals, making them crucial for imaging and detecting applications. LEDs and LDs work by inserting electrical power into a p-n junction, generating incoherent radiation in LEDs and coherent radiation in LDs, which are broadly used in demonstrations, communication, and illumination resolutions [1].

Solar cells are critical for green energy generation and sustainability, proposing a substitute to fossil fuels by altering sunlight into serviceable electricity. They are expansively used in solar power plants for broad-scale electricity generation, in astronomical technology for powering satellites, in consumer electronics such mobile chargers and watch, and in smart grids for efficient power storage and circulation. Their importance extends to agriculture, where they enable precision farming through automated arrangements, and in shipping, where they contribute to solar-driven vehicles and charging stations. The usefulness of a solar cell is largely determined by its transformation efficiency, power output, and solidity under environmentally friendly conditions [2-4].

High-quality PV cells should display definite characteristics such as high energy conversion efficiency, negligible recombination losses, and large carrier mobility for upgraded charge transport, durability and solidity to survive environmental dissimilarities, wide spectral absorption for operative photon deployment, and cost-effectiveness to confirm financial feasibility [2-5]. Numerous materials are commonly employed in marketable PV cells, each offering exclusive advantages and drawback. Silicon (Si) rules the market with PCE reaching 10.2-27.1%, while gallium arsenide (GaAs) achieves large PCE of 18.4-29.1% but remains costly, making it appropriate for space applications. Perovskite-based compounds offer 23.2-25.2% PCE but face stability limitation, whereas cadmium telluride (CdTe) provide an economical solution with 21% efficiency, although toxicity is a fear. Copper indium gallium selenide (CIGS) PV cells are lightweight and flexible, transporting efficiency 23.35%. Organic photovoltaics (OPV) are renowned for their flexibility and cost-effectiveness, but their PCE remains relatively low at 14.5-15.8%. Emerging graphene-based PV cells exhibit large conductivity and potential for commercialization, though they are still undergoing research and progress [6].

On the other hand, Photodetectors are essential for detecting optical signals in various wavelength regions, including ultraviolet (UV) (300-380 nm), visible light (380-780 nm), and infrared (NIR) (780-3000 nm). These devices convert optical signals into electrical signals, which can be used for real-time processing or stored for future applications [7]. The primary operational principle of photodetectors is the generation of photo-current upon light absorption. In recent years, photodetectors have found applications in diverse fields such as optical telecommunications for high-speed data transmission, biomedical imaging for diagnostics, electronic vision systems, epidermis sensors, oximeters, wearable devices, and foldable displays, chemical sensing, flame detection, ozone-hole analysis, missile plume sensing, and satellite imaging. They also contribute to night vision technology, spectroscopy, and astronomical studies, offering insights into high-radiation environments and deep-space observations [8-16]. A PD of high quality can be conceded by its direct band gap, which allows it to absorb light efficiently. It should have high sensitivity, and be worthy of absorbing light in different regions of the electromagnetic spectrum. It should also have a low dielectric constant, high signal-to-noise ratio (SNR), larger carrier mobility, high spectral selectivity, high transmittance, and high speed. Furthermore, it should be able to sense both high and low carrier concentrations and stay stable and hydrophilic with high responsivity and detectivity [15-18]. The most frequently used marketable photodetectors are: GaN, HgCdTe, Si, InSb, graphene, and InGaAs [18-19]. A $WS_2/AlO_x/Ge$ hetero-junction offers a responsivity (R) of 0.6345 A/W and detectivity ($D^*$) of $4.3\times10^{11}$ Jones. Similarly, a $PtSe_2$/Si-based PD provides an R of 0.52 A/W and $D^*$ of $3.26\times10^{13}$ Jones, while a $MoS_2$/Si-founded PD offers an R of 0.3 A/W and $D^*$ of $1\times10^{13}$ Jones. Some other instances of PD are $PtSe_2$/GaAs, $Bi_2Te_3$/Si and Graphene/Si-based provide the R of 0.262 $AW^{-1}$, 1 $AW^{-1}$, and 0.73 $AW^{-1}$ with the D* of $1\times10^{12}$ Jones, $2.5\times10^{11}$ Jones and $5.77\times10^{13}$ Jones respectively. Besides, experimental PDs such as Mg-$ZnSnP_2$ and Sn-$ZnSnP_2$ offer PD parameters in the range of 0.03 to 0.2 A/W and $1.62\times10^{11}$ to $4.62\times10^{12}$ Jones. Finally, a simulated PD based on PbS/$TiS_3$ offers the R of 0.36 A/W and the D of $3.90\times10^{9}$ Jones [19-26].

However, The PV cells and photosensors that rely upon graphene technology have several drawbacks, such as limited absorption capability, being toxic to the environment, being very expensive, having a zero tunable band gap, low carrier life, and also providing low efficiency responsivity and detectivity. Moreover, Si based SC-PD require complex and costly manufacturing processes, while GaAs SC-PD offer superior efficiency and PD performance but remain prohibitively expensive for large-scale applications. Perovskite SC-PD display excellent potential but suffer from degradation over time, affecting their longevity. CdTe and CIGS SC-PD provide competitive efficiency but introduce environmental concerns due to toxic material usage. Due to these critical issues, they are unsuitable for commercial use and human health care. They are also not suitable for military use in war [18-19, 25]. Herein, a novel $CdSnP_2$ (CTP) based double hetero-junction SC-PD has been proposed which contains high efficiency alongside high responsivity and detectivity along with different properties that a best-quality SC-PD owns.

The CTP compound has a chalcopyrite-type composition that includes Group II-IV-$V_2$ and I-42d Space group with a tetragonal structure. This compound has significant properties such as a low dielectric constant, lower noise rate, higher absorption coefficient, direct band gap (1.17 eV), better mobility, higher photosensitivity, low effective mass and dark condition resistivity of $10^4$-$10^5$ Ω/cm. Under light conditions, the resistivity is lower, about 3 to 4 times than in dark conditions. This means that the resistivity is low for n-type and high for p-type CTP. [27-39]. Moreover, CTP compound exhibits a high integral photoresponse, luminescent features and also shown improve radiative ability when doped with Cu or Ag [37].Those properties make CTP fit for usage in various fields, including Photo-Voltaic detectors (PVD), Solar fields, Light emitting diodes (LED), non-linear optics, oscillators, communication sector as an optical medium, and spintronic apparatus [30-31,34]. Additionally, stimulated emission of CTP compound crystal by double photon excitation shows that CTP can be also used a laser fabrication compound and the whole process is carried out by electron pumping method [38]. Considering the importance of CTP compound as a absorber layer different deposition techniques has been used to deposition thin film Such as Liquid epitaxy, photoelectrochemical and So on [36, 39]. In solvent method, CTP compound crystal hold a large amount of free charge carriers (>$10^{16}$ $cm^{-3}$) [37]. With these properties, CTP is an admirable appliance in SC-PD manufacturing, both for marketable and human welfare purposes.

To reduce surface recombination of the CTP layer, utilized a wide-bandgap compound has been as a window layer. At the upper end of the thick CTP layer, a thin window layer is conducted to create the p-n Junction. For a window layer to be effective, it must have certain properties such as a high band gap, low series resistance, and high transparency. CdS, which belongs to the II-VI group, has good quality window layer peculiarities like tunable wide band gap, chemical and physical stability, high-resolution transparency, and higher absorption ability in the UV region. [40-41]. Those superb properties of CdS make its application in different fields such as SC-PD, laser, transistor, LED, and so on [41].Bearing in mind the importance of CdS as a window layer, CdS deposited by various deposition methods such as physical vapor deposition (PVD), chemical bath deposition (CBD), molecular beam epitaxy (MBE), metal-organic chemical vapor deposition (MOCVD), close-spaced sublimation (CSS), electrodeposition, sputtering, pulsed laser deposition (PLD), spray pyrolysis and successive ionic layer adsorption and reaction (SILAR) [8]. For proper alignment and considering different properties and other factors, the CdS compound is widely used as a window layer [41].

A BSF layer is indispensable for the optimal performance of the CTP Solar cell and Photo-detector. It helps in bring down the dark current by built-in voltage and minimizing the recombination of minority charge carriers. This is achieved by forming an electric field between the CTP layer and the highly doped BSF layer, which develops the $J_{SC}$ and $V_{OC}$ [42]. The CGS layer, which belongs to the I-III-$VI_2$ family, can be used as a BSF layer owing to its direct energy gap, lofty absorption, and physical and chemical stability. The suitable band alignment and excellent electrical and optical properties of the I-III-$VI_2$ family compound i.e. CGS make it a first-rate back layer with CTP compound. Regarding the wide applicability in photonic, CGS material thin film are deposited by several fabrication method such as pulsed electron deposition, thermal co-evaporation, molecular beam epi-taxy (MBE), pulsed laser deposition (PLD), metal-organic chemical vapor deposition (MOCVD) [43]. In fact, CGS has been previously utilized as a back layer to boost the appliance's performance. [44-45].

Recently, machine learning (ML)-assisted optimization has emerged as a powerful approach for accelerating photovoltaic device engineering by identifying complex nonlinear relationships between material properties and device performance [46]. However, despite significant progress in ML-guided solar-cell optimization, the integration of ML/DL techniques with SC-PD architectures remain largely unexplored. To the best of our knowledge, no previous study has reported a hybrid SCAPS-ML framework for SC-PD device optimization together with explainable AI-based design-rule extraction. In this work, multiple ML/DL models combined with SHAP analysis are employed to predict high-performance SC-PD configurations and establish physics-guided optimization strategies through band-offset engineering and material-property analysis. This data-driven framework not only accelerates device optimization but also provides interpretable insights for the intelligent design of high-performance SC-PD devices.

This article aims to explore a highly efficiency alongside responsivity and detectivity SC-PD which is predicted by multiple ML/DL models combined with SHAP analysis. The predicted best $CdSnP_2$ SC-PD structure with CGS as a BSF layer, along with a CdS buffer layer and Al-Ni metal contacts has regulated to determine the optimized condition of each layer by SCAPS-1D Simulator software. This articles also studied the influence of temperature, series, and shunt resistance on SC-PD performance parameters on the proposed structure.

## 2. Research Approach and Mockup Parameters

### 2.1 Simulation Approach and Modeling Factors

Figure 1(a) exhibit a Graphic diagram and 1(b) an electronic energy diagram of the CTP SC-PD with an energy gap (EG) of 1.17 eV makes a hetero-junction with CdS owns EG of 2.4 eV acts as the electron transporting layer (ETL) layer. The CGS with EG of 1.66 eV is a trust-worthy aspirant to form a p-$p_+$ connection with the CTP and Al and Ni are employed as a Cathode and Anode with the work function (WF) of 4.2 and 5.25 eV [29,40,45].Al/n-CdS/p-$CdSnP_2$ /$p^+$-$CuGeSe_2$ /Ni DH SC-PD has statistically determined its $J_{SC}$, $V_{OC}$, FF, PCE, Responsivity, and detectivity by shifting corporeal parameters such as thickness, doping concentration, and defect density. This structure has been arithmetically examined by SCAPS-1D which a group of experts developed at the University of Gent, Belgium [40].

SCAPS-1D emulator software system regulates the different PV parameters from the J-V section and PD parameters from the QE section by resolving the Poisson equation [30]. Below all the equations associated with SCAPS are shown where equations (1),(2), and (3) are the Poisson, hole, and Electron continuity equations respectively [41].

$$\frac{\varepsilon}{e}\frac{\partial^2\Psi}{\partial y^2} = \left[n(y) - p(y) + Ni_A - Ni_D + \rho_n - \rho_p\right] \quad (1)$$

$$\frac{1}{e}\frac{\partial J_p}{\partial y} = C_G - R_R(y) \quad (2)$$

$$\frac{1}{e}\frac{\partial J_n}{\partial y} = -C_G + R_R(y) \tag{3}$$

In equation (1), the relative permittivity, electrostatic potential, electron charge, the concentration of ionized acceptors and donors, charge carriers (electron€-hole(p)) density, and distribution of e-p are represented emblematically by Ψ,ε, e, $Ni_{A,}$ and $Ni_D$, , p,n $\rho_p$, and $\rho_n$. In equations (2) and (3), $J_n$ , $J_p$ , $C_G$, $R_R$ emblematic representation of e-p current density, overall carrier generation, and the recombination percentage. Drift-diffusion equation of holes and electrons is shown below which it considered a transpiration property [45].

$$J_p = -\frac{\mu_p p}{q}\frac{\partial E_{F_p}}{\partial y} \tag{4}$$

$$J_n = -\frac{\mu_n}{q}\frac{\partial E_{F_n}}{\partial y} \tag{5}$$

In equations (4) and (5), the hole and electron mobility, the Fermi points of extrinsic compound charge transporters are symbolically represented by $\mu_p$, $\mu_n$ , $E_{Fp,}$ and $E_{Fn}$ .

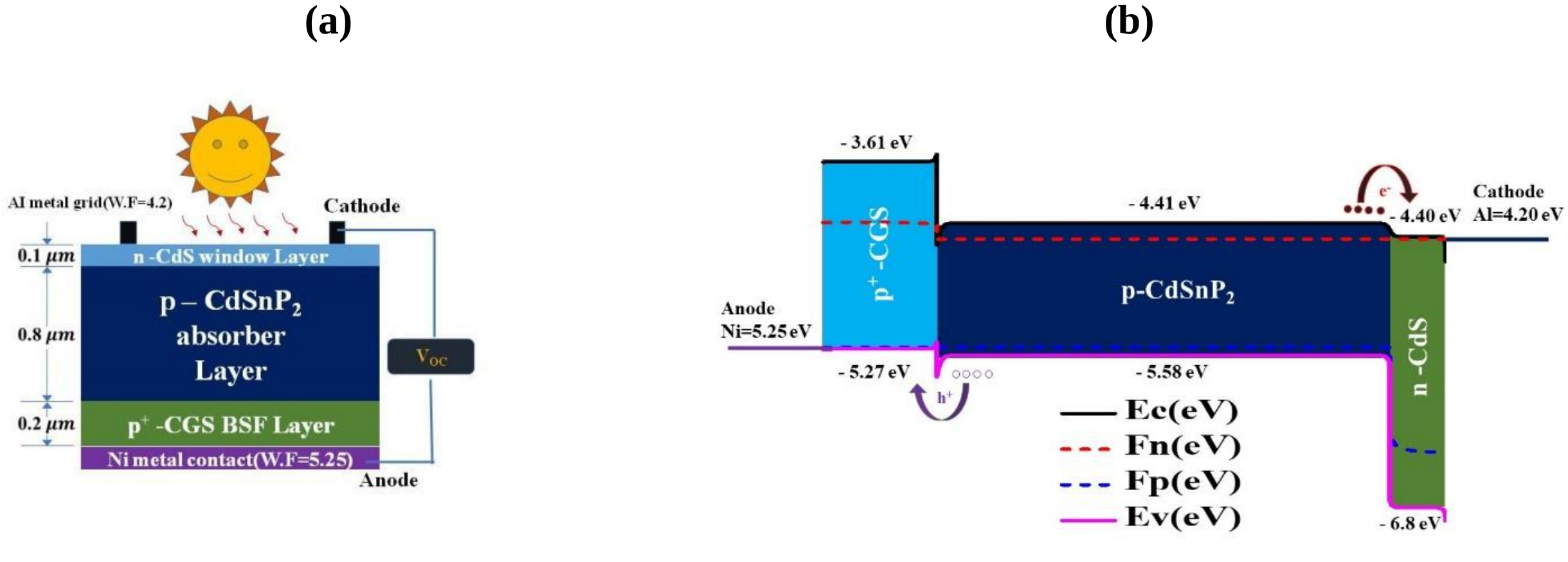

**Figure 1:** (a) Graphic (b) Electronic energy gap drawing of n-CdS/p-$CdSnP_2$/$p^+$-$CuGeSe_2$ Photodetector.

For $CdSnP_2$ Photodetector PV and PD parameters calculation, consider appliance temperature 300 K and the SC-PD appliance illumined under Solar radiation of Air Mass (AM) 1.5 G with Power intensity 1000 W/m$^2$. The series and Shunt resistance are considered ideal cases. The interface defect of p-$CdSnP_2$ /n-CdS and p-$CdSnP_2$ /p$^+$-CGS interface are consider as $1.0\times10^{11}$ cm$^{-2}$**.** All other physical parameters for each layer are shown in Table 1.

**Table 1:** Collected parameters from the published article for imitating n-CdS/p-$CdSnP_2$/p$^+$-CGS SC-PD.

| **Parameters** | ***n*-CdS [40]** | ***p*-$CdSnP_2$ [27,35,47]** | ***p*$^+$- CGS [45]** |
|---|---|---|---|
| Layer | Window | Absorber | BSF |
| Depth (μm) | 0.100 | 0.8 | 0.200 |
| Energy gap, $E_g$ [eV] | 2.4 | 1.17 | 1.660 |
| Affinity of electron, $\chi$ [eV] | 4.400 | 4.410 | 3.610 |
| Relative permittivity | 10 | 11.8 | 10.900 |
| Operative DOS at CB [c/m$^3$] | $2.2\times10^{18}$ | $3.33\times10^{17}$ | $2.20\times10^{17}$ |
| Operative DOS at VB [c/m$^3$] | $1.8\times10^{19}$ | $1.44\times10^{19}$ | $1.8\times10^{18}$ |
| Thermal velocity of Electron (cms$^{-1}$) | $1.0\times10^{7}$ | $1.0\times10^{7}$ | $1\times10^{7}$ |
| Thermal velocity of Hole (cms$^{-1}$) | $1.0\times10^{7}$ | $1.0\times10^{7}$ | $1\times10^{7}$ |
| Electron Mobility $\mu_n$ [cm$^2$V$^{-1}$s $^{-1}$] | $1.0\times10^{2}$ | $2.0\times10^{3}$ | $1.0\times10^{2}$ |
| Hole mobility $\mu_p$ [cm$^2$V$^{-1}$s$^{-1}$] | $2.5\times10^{1}$ | $1.0\times10^{2}$ | $2.5\times10^{2}$ |
| Donor concentration [c/m$^3$] | $1.0\times10^{18}$ | 0 | 0 |
| Acceptor concentration [c/m$^3$] | 0 | $1.0\times10^{18}$ | $1.0\times10^{18}$ |
| Total defect density, $N_t$ [cm$^{-3}$] | $1.0\times10^{14}$ | $1.0\times10^{14}$ | $1.0\times10^{14}$ |

### 2.2 Machine Learning Methodology and Data Acquisition

The ML/DL models has been implemented using Python 3.11. The complete ML/DL-based optimization framework for the SC-PD devices has been demonstrated in figure 2. A total 13 different configurations of solar cell-photodetector (SC-PD) structure dataset has been incorporated to the ML framework. The dataset used for training and evaluation was engendered through replications performed in the SCAPS-1D environment by merging multiple window layer and back surface field (BSF) materials. The window layer materials included CdSe, CdS and for the BSF layer, the materials included AlSb, $BaSi_2$, CFTS, CGS, SnS, $WSe_2$, GeS. For the input features of the ML models, both physical and electronic properties of the dual heterostructure (SC-PD) were considered. The window layer features included: material type, bandgap, and electron affinity. For the absorber layer: the features included width, bandgap, electron affinity, dielectric permittivity, effective DOS at the CB and VB, electron and hole thermal velocity, electron and hole mobility, acceptor density, and defect density. For the BSF layer: material type, bandgap, and electron affinity were used as input parameters. Additionally, CB and VB offsets at window layer/absorber layer (CB_offset1, VB_offset1) and absorber/BSF layer (CB_offset2, VB_offset2) has been used as input parameters. To ensure numerical stability and improve model learning efficiency, a logarithmic (log10) transformation was applied to all high-magnitude input features surpassing $10^4$, as well as to the output parameter detectivity $D^*$. This log transform converts extremely large values (e.g., $10^{15}$ - $10^{21}$) into a manageable scale (15–21), that helps reduce skewness and variance in the dataset. This transformation improves the ability of the ML models to identify underlying relationships more effectively. Finally, all predictions for $D^*$ were inverse-transformed using $10^x$ to restore physically meaningful values in the original scale.

By synthetically changing these input features, a total of 3625 data sample were generated using the SCAPS-1D across all 13 structural configurations. In this dataset 11 structure data (3096 data points) was used for training purpose and rest 2 structure data (529 data points) was used for validation of the ML models. This strictly structure wise based splitting causes ML model to predict completely unknow structure behaviors. From this splitting technique approximately 85% of the data sample was used for training purpose and 15% was used for testing the ML/DL models.

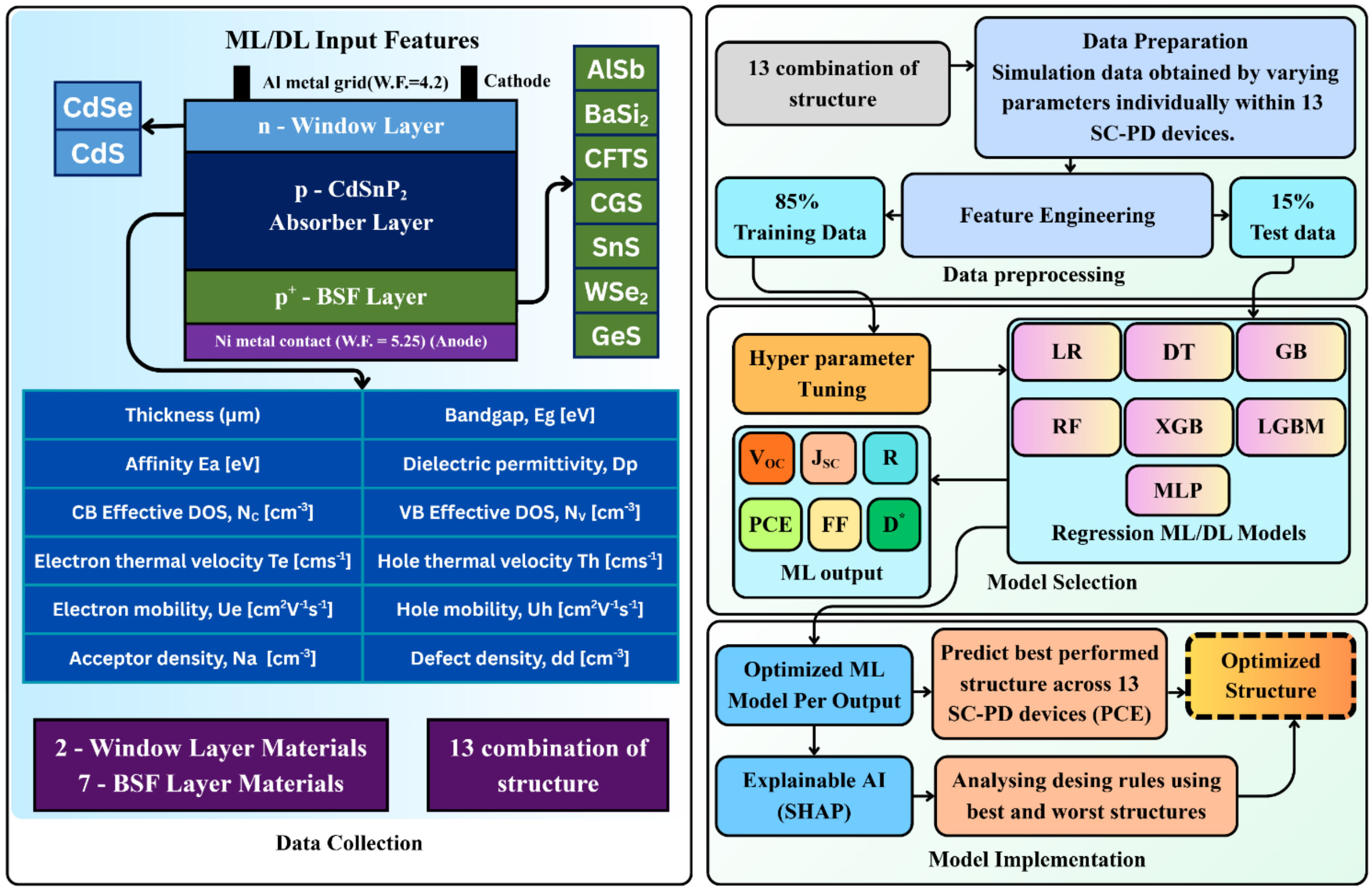

**Figure 2:** Workflow illustrating the ML/DL-based optimization strategy for designing high-performance SC-PD devices.

### 2.2.1 Machine Learning and Deep Learning Algorithms

In this study, six machine learning (ML) and one deep learning (DL) algorithm have been implemented, as aforementioned, to model the relationship between input parameters and target outputs. The selected models included linear regression, decision tree, random forest, gradient boosting, extreme gradient boosting, light gradient boosting machine, and a multi-layer perceptron (MLP).

Among these models, ensemble-based methods including RF, GB, XGB and LGBM are utilized to understand the complex nonlinear relationships and higher-order feature interactions. This ability enables the identification of highly efficient structural configurations and supports effective optimization, which overcomes the conventional analysis limitations, thereby accelerating the overall research and development process in material science [48]. The MLP a deep learning model further extends this capability by learning hierarchical feature representations through multiple fully connected layers and nonlinear activation functions (like ReLu). [49]

For hyperparameter tuning of the models to improve model performance, GridSearchCV has been used in this study with a predefined hyperparameter combination. The optimal settings were used for each model individually. To better understand the model behavior, SHAP was applied to the best performed model to break down each prediction into contributions from individual characteristics relative to a reference line. This offers a translucent view of how features influence the output and interact with one another [50].

## 3. Result and discussion

### 3.1 ML structure for optimal framework Identification

The dataset for ML model includes 13 SCAPS-1D simulated SC-PD device structures, with approximately ~300 samples per structure, resulting in a total of 3626 data samples with 22 input features. The input parameter ranges for Window, Absorber and BSF layer are summarized in Table 2. The hyperparameter has been utilized using the GridSearchCV with 5-fold cross justification from Kfold. In addition, this structure-wise splitting is used for the validation, where two entire unseen device structures are kept for testing and the lasting eleven are utilized for training. This distribution strategy ensures higher validation approach to model performance.

**Table 2:** Input parameter ranges varied across the 13 SC-PD device structures used for ML DL models.

| **Window Layer** | **Absorber Layer** | | | **BSF Layer** |
|---|---|---|---|---|
| CdSe | **Parameters** | **Ranges** | **Steps** | AlSb |
| | Layer Width (μm) | 0.1 – 2.5 | 25 | |
| | Bandgap, $E_g$ [eV] | 0.6 – 1.8 | 25 | $BaSi_2$ |
| | Affinity $E_a$ [eV] | 3.7 – 5.06 | 25 | |
| | Dielectric permittivity, $D_p$ | 3 – 27 | 25 | CFTS |
| | CB Effective DOS, $N_C$ [$cm^{-3}$] | $1\times10^{15}$ - $6.92\times10^{18}$ | 25 (log) | |
| CdS | VB Effective DOS, $N_V$ [$cm^{-3}$] | $1\times10^{18}$ - $1\times10^{20}$ | 25 (log) | CGS |
| | Electron thermal velocity, Te [$cms^{-1}$] | $5\times10^{6}$ - $2\times10^{9}$ | 25 | |
| | Hole thermal velocity, Th [$cms^{-1}$] | $5\times10^{6}$ - $2\times10^{9}$ | 25 | SnS |
| | Electron mobility, $U_e$ [$cm^2V^{-1}s^{-1}$] | 800 - 3200 | 25 | |
| | Hole mobility, $U_h$ [$cm^2V^{-1}s^{-1}$] | 10 - 250 | 25 | $WSe_2$ |

|  | Acceptor density, $N_a$ [$cm^{-3}$] | $1\times10^{16}$ - $1\times10^{20}$ | 25 (log) |  |
|---|---|---|---|---|
|  | Defect concentration, dd [$c/m^3$] | $1\times10^{12}$ -$1\times10^{16}$ | 25 (log) | GeS |

### 3.1.1 Correlation investigation Between Input parameters and Target Variables

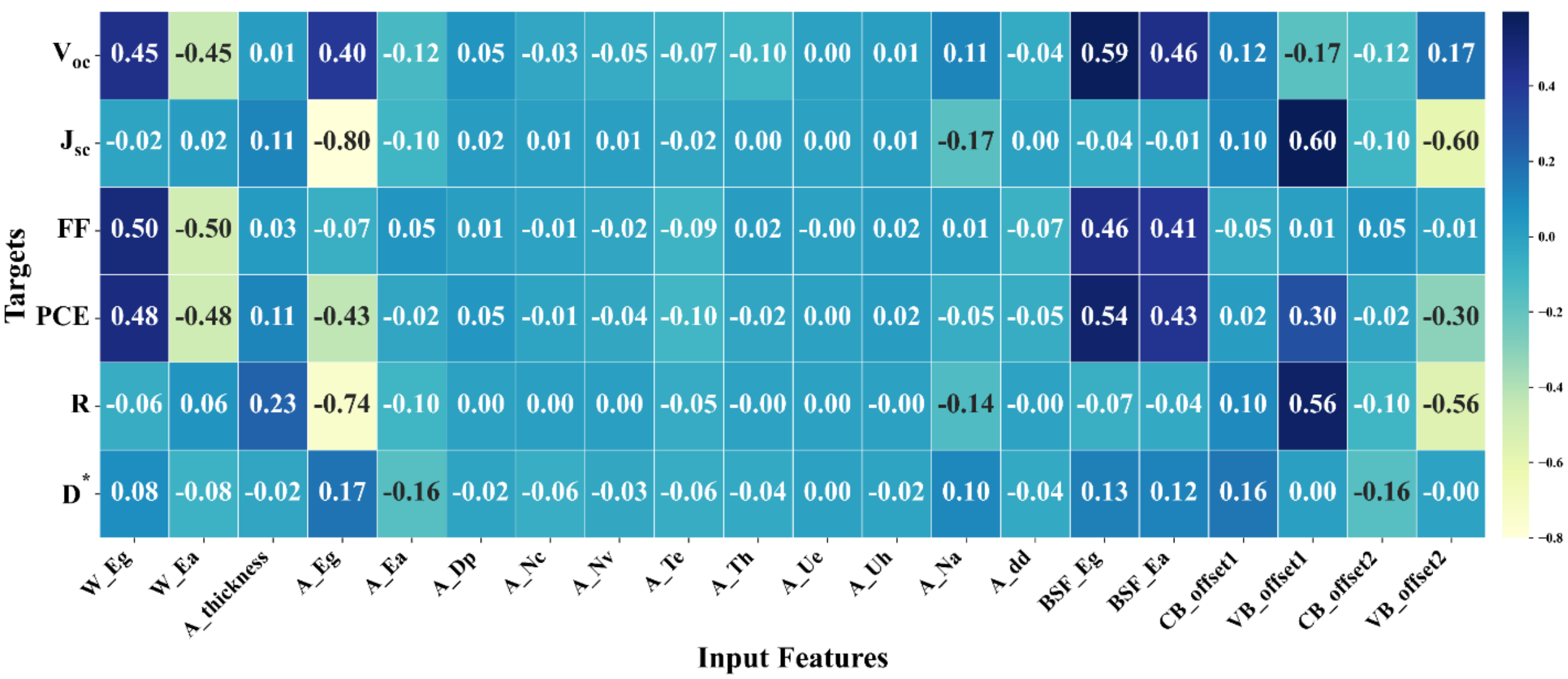

**Figure 3:** Average correlation matrix depicting the relationships between input features and target parameters across 13 SC-PD device structures.

Figure 3 shows the average correlation matrix illustrating the influence of input features for individual target parameters across 13 SC-PD structures. The correlation was computed for individual structure-data and then averaged across all structures. However, for the window and BSF layer parameters (W_$E_g$, W_$E_a$, BSF_$E_g$, BSF_$E_a$), the correlation was calculated using the entire dataset because the values remained constant within individual structure-data but varied across different structures. The figure reveals that $V_{OC}$ has strong positive correlation with the W_$E_g$, A_$E_g$ and BSF_$E_g$, BSF_$E_a$. It also exhibits positive correlations with A_$N_a$, CB_offset1 and VB_offset2. However, the W_$E_a$ and VB_offset1 show strong negative influence with $V_{OC}$. For the output $J_{SC}$, a strong negative correlation is observed for A_$E_g$ and VB_offset2 parameter.

$J_{SC}$ shows positive correlations with A_thickness and VB_offset1, whereas A_$N_a$, A_$E_a$, CB_offset2 demonstrate negative correlation. The figure further reveals that FF is positively correlated with the W_$E_g$, BSF_$E_g$ and BSF_$E_a$, while W_$E_a$ exhibits a negative correlation. For PCE, strong positive correlations are observed with the W_$E_g$, BSF_$E_g$, BSF_$E_a$ and VB_offset1. Conversely, W_$E_a$, A_$E_g$ and VB_offset2 exhibit strong negative correlations. The responsivity R shows a strong positive correlation with A_thickness and VB_offset1. However, A_$E_g$, A_$N_a$, and VB_offset2 demonstrate negative correlations. The detectivity $D^*$ displays positive correlations with A_$E_g$, A_$N_a$, BSF_$E_g$, BSF_$E_a$, CB_offset1. The A_$E_a$ and CB_offset2 parameters are negatively correlated as observed in the heatmap for detectivity. Band offset engineering directly affects the interfacial carrier transport and recombination which play vital roles in determining optimal device performance [51]. The high correlations observed in the figure further confirms the strong dependence of device outputs on band offset alignment.

### 3.1.2 Parameter for Model Performance

In ML/DL investigation and modeling, three valuation parameters— (MAE, RMSE, $R^2$ score) are employed to comprehensively assess model performance and capture different aspects of prediction error. In ML/DL performance analysis, relying on a single metric is often insufficient to represent the overall behavior of a model, so, the use of multiple complementary metrics enables a more robust and balanced evaluation [52]. In this study, these three metrics were utilized to deliver a complete performance analysis of the models.

### Optimization of Hyperparameters

Table 3 shows the adjusted hyper-parameter standards for each individual model of the dataset. The hyper-parameter optimization was done through using the GridSearchCV. The finest performance metrics was achieved for each ML/DL model by using this optimal parameter values, in which the RF model shows the best prediction for the target $V_{OC}$, $J_{SC}$ and $D^*$, XGB model performed best for FF and R, finally for the PCE, GB model displays the most operative exploitation of the dataset.

**Table 3:** Finest Hyperparameter Standards for Diverse ML Mockups.

| Machine Learning Model | Hyper-parameter Space | Optimal Parameter |
| --- | --- | --- |
| RF | n_estimators: [100, 200, 300, 400], max_depth: [10, 15, 20] | n_estimators: 400, max_depth: 20 |
| GB | learning_rate: [0.01, 0.05, 0.1], max_depth: [2, 3, 4, 5, 6], n_estimators: [200, 300] | learning_rate: 0.1, max_depth: 5, n_estimators: 500 |
| LR | - | - |
| DT | max_depth: [3, 5, 10], min_samples_split: [2, 5, 10], min_samples_leaf: [1, 2, 4] | max_depth: 10, min_samples_leaf: 1, min_samples_split: 2 |
| XGB | n_estimators: [300, 500], learning_rate: [0.03, 0.05, 0.1], max_depth: [3, 4, 5, 6], subsample: [0.8, 1.0], colsample_bytree: [0.8, 1.0] | n_estimators: 500, learning_rate: 0.1 max_depth: 5, subsample: 1, colsample_bytree: 1 |
| LGB | learning_rate: [0.01, 0.05, 0.1], max_depth: [4, 5, 6], n_estimators: [100, 300, 500], | learning_rate: 0.1, max_depth: 6, n_estimators: 500, |
| MLP | batch_size: 16, architecture: [64-32-6 (Dense layers)], activation: relu, output_activation: linear, optimizer: adam, learning_rate: 0.001, loss: mse, epochs: 100, early_stopping_patience: 10, regularization: L2(0.001), dropout: [0.3, 0.2], callbacks: EarlyStopping, ReduceLROnPlateau | |

The overall evaluation metrics of seven models for six target variables are shown in Table 4. The table shows that for five Target outputs the RF model performed extremely well compared to others. Specifically, the model achieved the highest for predicting $V_{OC}$, $J_{SC}$, PCE, R and $D^*$ with a $R^2$ score of 0.9947, 0.9928, 0.9913, 0.9918 and 0.9942. The model also has the lowest MAE and RMSE score for this Targets. This revels the multiple tree-based model has higher chance to capture the nonlinear relationship of these three outputs. On the other hand, the GB model has the highest $R^2$ score of 0.9784 for predicting PCE. The XGB and GB model has almost similar $R^2$ score for PCE prediction, but the RMSE value is lowest for the GB model (1.091), indicating its superior predictive performance for SC-PD device efficiency within the 13-structure dataset.

The LR model has the worst performance among all of the other models, which describes the complex, non-linear correlation between the features and targets. For the prediction of FF and R, the XGB model has the highest $R^2$ score (0.9576, 0.9832) compared to other models, along with the lowest MAE and RMSE score. Overall, the results demonstrate that all the tree-based ensemble models- including RF, GB, XGB, LGB are powerful and effective to capture the complex, nonlinear relationships. In contrast, the single tree-based (DT) model showed comparatively lower performance across most targets.

The deep learning model MLP also have a decent $R^2$ score for all output targets except the FF. Figure 4 shows the training and validation performance of the MLP model combinedly across all six-output target for (a) MSE and (b) Mean Absolute Error (MAE) over training epochs. It is seen that both MSE and MAE decrease gradually with increasing epochs, which indicates effective learning and convergence of the model.

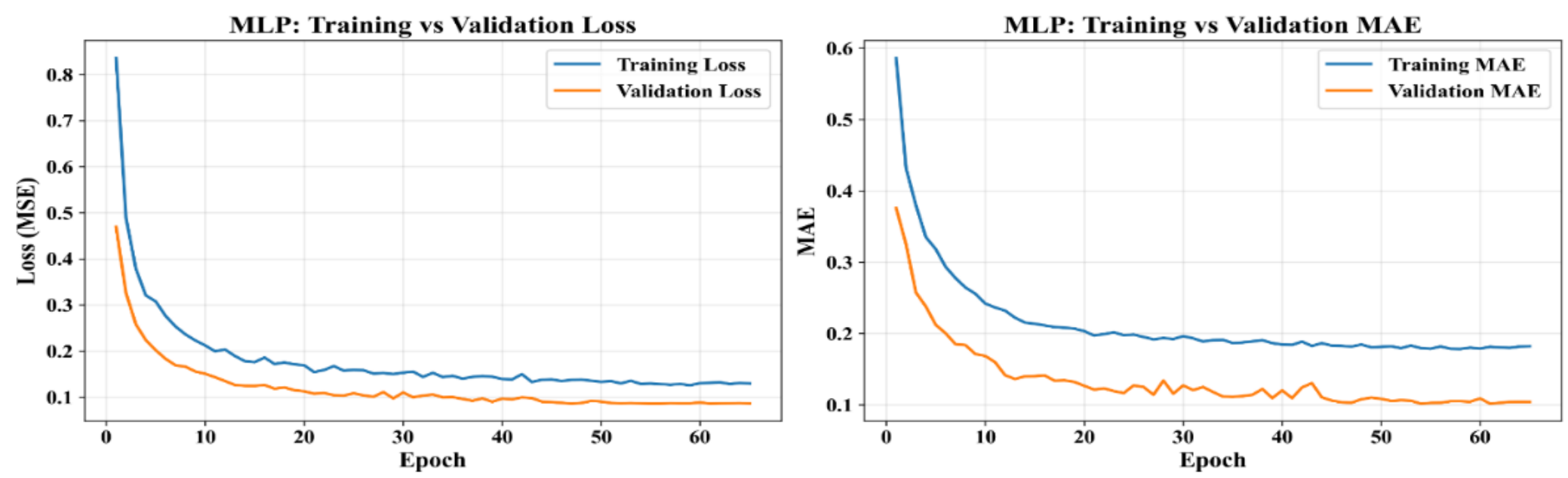

**Figure 4:** Training and validation loss of the MLP model showing aggregated MSE and Mean Absolute Error (MAE) across all six output targets over training epochs.

**Table 4:** Comparison of ML/DL Model Performance Metrics (test set).

| Model | Metric | Voc | Jsc | FF | PCE | R | $D^*$ |
|---|---|---|---|---|---|---|---|
| **RF** | MAE | 0.0128 | 0.3107 | 0.6121 | 0.5827 | 0.004 | 0.11 |
| | RMSE | 0.0144 | 0.4294 | 1.4054 | 0.7261 | 0.0092 | 0.1251 |
| | **$R^2$ Score** | **0.9947** | **0.9928** | 0.9521 | **0.9913** | **0.9918** | **0.9942** |
| **GB** | MAE | 0.0187 | 0.5237 | 0.7987 | 1.0309 | 0.0025 | 0.1545 |
| | RMSE | 0.0207 | 0.6388 | 1.2703 | 1.2168 | 0.0123 | 0.1736 |
| | **$R^2$ Score** | 0.989 | 0.9841 | **0.9609** | 0.9755 | 0.9852 | 0.9889 |
| **LR** | MAE | 0.271 | 0.2773 | 0.4988 | 0.3786 | 0.2718 | 0.2718 |
| | RMSE | 0.4153 | 0.5741 | 0.6913 | 0.5496 | 0.5716 | 0.4212 |
| | **$R^2$ Score** | 0.8184 | 0.6355 | -3.4262 | 0.5589 | 0.6378 | 0.8137 |
| **DT** | MAE | 0.0162 | 0.5379 | 0.7096 | 1.1038 | 0.0052 | 0.1409 |
| | RMSE | 0.0205 | 1.1529 | 1.4048 | 1.5936 | 0.015 | 0.1935 |
| | **$R^2$ Score** | 0.9892 | 0.9481 | 0.9522 | 0.9579 | 0.9781 | 0.9862 |
| **XGB** | MAE | 0.0211 | 0.5155 | 0.8277 | 0.8708 | 0.0032 | 0.1533 |
| | RMSE | 0.0239 | 0.6575 | 1.5371 | 1.0362 | 0.0103 | 0.1733 |
| | **$R^2$ Score** | 0.9854 | 0.9831 | 0.9427 | 0.9822 | 0.9897 | 0.9889 |
| | MAE | 0.0189 | 0.6519 | 1.2651 | 0.881 | 0.0084 | 0.1464 |

| LGB | RMSE | 0.0251 | 1.1687 | 2.6096 | 1.2366 | 0.027 | 0.1992 |
|---|---|---|---|---|---|---|---|
| | **$R^2$ Score** | 0.9839 | 0.9467 | 0.8349 | 0.9746 | 0.9292 | 0.9854 |
| **MLP** | MAE | 0.0176 | 0.4879 | 2.4186 | 0.7245 | 0.0119 | 0.1478 |
| | RMSE | 0.0283 | 1.382 | 3.6394 | 1.2554 | 0.0282 | 0.2415 |
| | **$R^2$ Score** | 0.9795 | 0.9254 | 0.6789 | 0.9739 | 0.9229 | 0.9785 |

Table 5 shows the best performing model for each output target, with both training and test set metrics. The best model is selected based on the highest Test $R^2$ score.

**Table 5:** Performance metrics of best ML models for individual output targets.

| Target | Best Model | Train MAE | Test MAE | Train RMSE | Test RMSE | Train $R^2$ | Test $R^2$ |
|---|---|---|---|---|---|---|---|
| $V_{OC}$ | **RF** | 0.0022 | 0.0128 | 0.0041 | 0.0144 | 0.99959 | 0.9947 |
| $J_{SC}$ | **RF** | 0.1003 | 0.3107 | 0.3683 | 0.4294 | 0.99521 | 0.9928 |
| FF | **GB** | 0.081 | 0.7987 | 0.1351 | 1.2703 | 0.99995 | 0.9609 |
| PCE | **RF** | 0.1476 | 0.5827 | 0.2977 | 0.7261 | 0.99899 | 0.9913 |
| R | **RF** | 0.0021 | 0.004 | 0.0071 | 0.0092 | 0.99559 | 0.9918 |
| $D^*$ | **RF** | 0.0201 | 0.11 | 0.0376 | 0.1251 | 0.99951 | 0.9942 |

Figure 5 represents the prediction vs actual values for all six targets using the best performed ML models. The close alignment of training and test data points along dashed line indicates strong predictive accuracy and generalization across all targets.

**(a)** **(b)** **(c)**

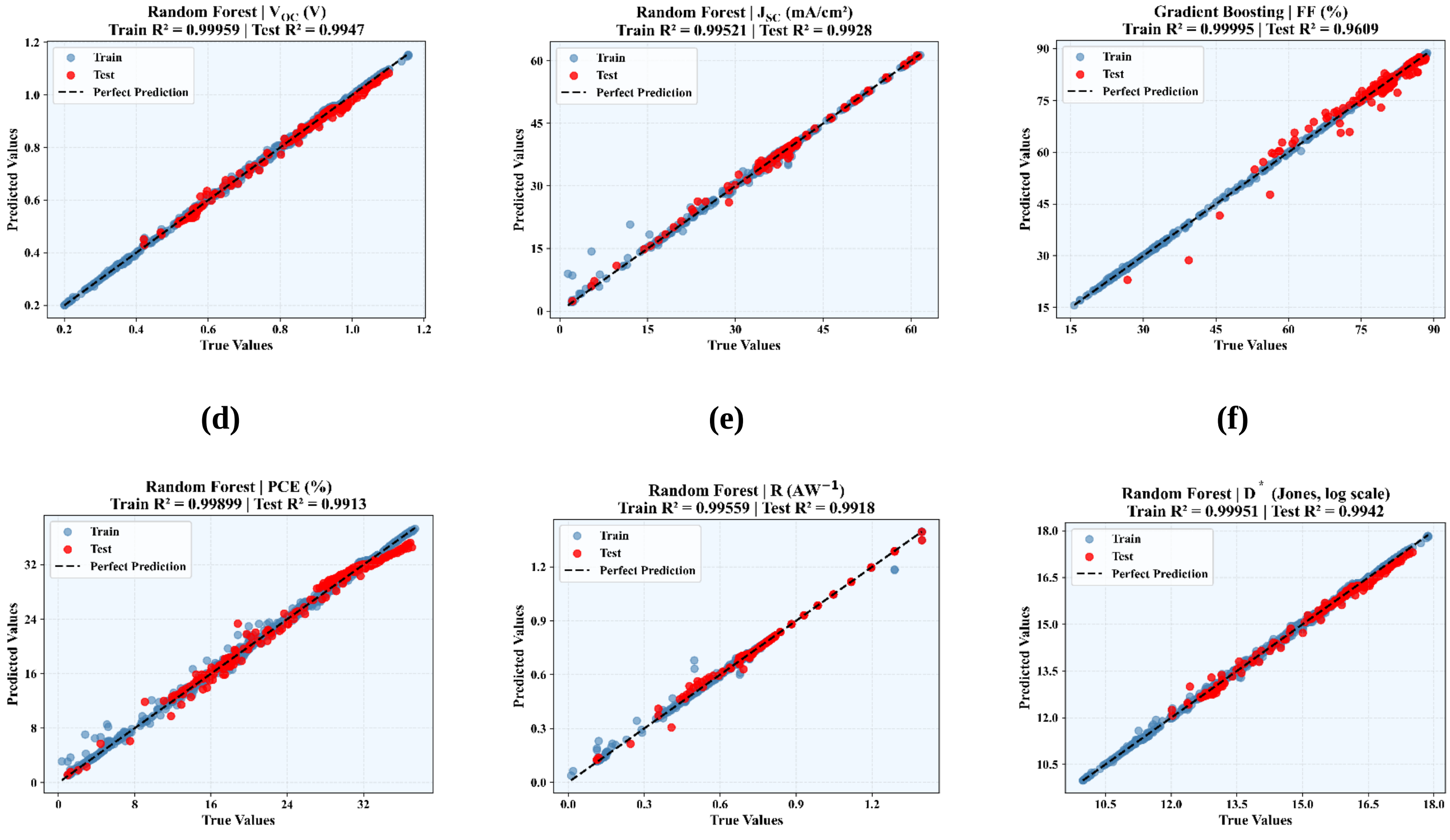

**Figure 5:** Predicted vs. true values for six output targets using the best-performing machine learning models. (a) RF for $V_{OC}$ (V) , (b) RF for $J_{SC}$, (c) XGB for FF, (d) GB for PCE, (e) XGB for R, and (f) RF for $D^*$ (log scale). The dashed line represents perfect prediction.

### 3.1.3 Machine learning approach to predict optimized structure

The best performed model has been utilized to predict the optimal structure of SC-PD devices. In this study, the best performed structure is selected based on PCE that has been predicted by the RF model. Consecutively, the model has been trained according to the same data-splitting strategy, and the top-ranked structure predicted by the optimized model exactly matches with the five highest-performing structures identified through SCAPS-1D simulations, showing only a minimal margin of error. The best SC-PD device structure predicted by the GB is shown in Table 6. The table indicates that the highest performing SC-PD device structure predicted by the RF model is CdS-CdSnP2-CGS, which also corresponds to the top-ranked device obtained from the actual simulation results.

**Table 6:** Comparison of Actual vs Predicted PCE Values (RF Model).

| Structure Name | Actual PCE (%) | Predicted PCE (%) | Error PCE |
|---|---|---|---|
| n-CdS-p-$CdSnP_2$-p+-CGS | 32.6906 | 32.6732 | 0.0174 |
| n-CdSe-p-$CdSnP_2$-p+-$WSe_2$ | 32.4826 | 32.4507 | 0.0319 |
| n-CdS-p-$CdSnP_2$-p+-$WSe_2$ | 32.4826 | 32.0760 | 0.4066 |
| n-CdS-p-$CdSnP_2$-p+-SnS | 29.9788 | 29.9585 | 0.0203 |
| n-CdS-p-$CdSnP_2$-p+-AlSb | 28.2122 | 28.1453 | 0.0669 |

### 3.1.4 Feature Impact analysis determined by the Models

#### 3.1.4.1 explainable AI (SHAP) Observation

The SHAP waterfall plots showing individual feature influence in figure 6 provide a quantitative explanation of how the most influential physical parameters impact the single absorber (Window-absorber-BSF) architecture across the SC-PD device performance (PCE, R, and $D^*$). For the high-PCE structure (Figure 6a), the baseline PCE has increased from 20.93% to the optimum value of 32.07%. The figure shows that the predicted output increases significantly from the baseline value to the optimized output, mostly driven by VB_offset2 (0.23eV), which exhibits the strongest positive contribution (+5.59%). This indicates the key impactful parameter, which highlights the critical role of proper valence-band alignment at the absorber/BSF interface which enables efficient hole extraction and reducing recombination losses. The CB_offset1 (-0.01 eV) also portrays a large positive impact (+2.36%). This directs the favorable conduction-band alignment further enhances carrier selectivity. The BSF layer ($WSe_2$) material also plays a key role (+2.05%) in predicting the PCE. Other positive contributions arise from VB_offset1 (1.22 eV, +0.63%) and CB_offset2 (0.71 eV, +0.29%). This confirms the optimization of both CB and VB offsets are essential for maximizing the efficiency of the proposed device. In opposition, BSF_$E_a$ (3.70 eV, −0.15%) and BSF_$E_g$ (1.65 eV, −0.09%) reveals slight negative contribution, which suggest that BSF electronic properties can slightly hinder PCE. Overall, this enhancement of the device performance reveals that band alignment and BSF engineering dominate the high-efficiency regime [51].

For the low-PCE structure (Figure 6b), the baseline PCE has reduced from 20.93% to the degraded value of 17.25%. The predicted efficiency has decreased due to adverse parameter configurations. The major negative effect is due to VB_offset2 (0.98 eV, −10.01%), which indicates severe misalignment that creates a large barrier for carrier transport and significantly increases recombination. Although CB_offset1 (-0.21 eV) compensates this reduction in performance since contributes positively (+6.17%) and it is insufficient to overcome the dominant negative impact. Other contributing factors include BSF material ($BaSi_2$, −0.39%), VB_offset1 (1.02 eV, +0.20%), and absorber doping A_Na ($1x10^{18}$, +0.18%), which provide only minor changes to the overall prediction. These results summarize that improper band alignment, particularly at the valence band, is the primary factor limiting PCE in low-performing structures.

In the case of responsivity R (Figure 6c and 6d), the SHAP contributions are comparatively smaller, as the value for R lies in range of ~0.4 $AW^{-1}$ to 0.7 $AW^{-1}$ [53-54]. For high responsivity structure (Figure 6c), the baseline R has increased from 0.712 $AW^{-1}$ to the optimum value of 0.716 $AW^{-1}$. For this structure, absorber bandgap A_$E_g$ (1.17 eV, −0.01 $AW^{-1}$) and VB_offset1 (1.22 eV,

+0.01 $AW^{-1}$) appears as the leading contributors in the model prediction. The BSF material ($WSe_2$), CB_offset1 (-0.01 eV) and VB_offset2 (0.23eV) also displays positive impact to the model prediction. However, the BSF_$E_g$ (1.65 eV) has a slight negative impact for the high R structure.

For the low-R structure (Figure 6d), the main factor is absorber layer thickness (A_thickness) (0.10 μm, −0.31 $AW^{-1}$), which significantly decreases R when the absorber thickness is reduced. Secondary contributions from BSF_$E_g$ (1.65 eV, −0.01 $AW^{-1}$), A_$E_g$ (1.17 eV, −0.01 $AW^{-1}$) shows minor negative impact and most other parameters exhibits a very small impact for predicting R. This suggests that thickness optimization plays a central role in controlling the responsivity of the device structure.

The model revealed some stronger impact on key parameters for predicting the detectivity $D^*$ (Log scaled) (Figures 6e and 6f). The dominant influence comes from BSF_$E_a$ (3.70 eV, +0.69 Jones) in the high-D structure (Figure 6e), revealing that proper value of the electron affinity at the back surface significantly improves carrier collection and reduces recombination. Subsequently, BSF layer ($WSe_2$, +0.38 Jones) and CB_offset1 (-0.01 eV, +0.34 Jones) directs the importance of both back-surface engineering and conduction-band alignment in predicting the detectivity. Additionally, positive effects from the W_$E_a$ (4.40 eV, +0.20 Jones) and W_$E_g$ (2.40 eV, +0.12) highlights the key role of the window layer in optimizing carrier transport and minimizing front-surface losses. On the other hand, CB_offset2, Window layer material, VB_offset2 also exhibited small increasing contribution of (+0.12), (+0.11) and (+0.1) Jones and a small negative influence from VB_offset1 (−0.03) suggests slight inefficiencies due to imperfect band alignment.

In contrast, the low-D structure (Figure 6f) is mostly influenced by a strong negative contribution from the same key parameter BSF_$E_a$ (3.30 eV, −1.89 Jones). This indicates that improper BSF electron affinity, which causes improper band-alignment, severely reduces detectivity. This reduction is partially compensated by positive contributions from W_Eₐ (+0.27), CB_offset1 (+0.19), and window layer properties (+0.16). Though these improvements are insufficient to counterbalance the major loss of detectivity. Further reduction in $D^*$ comes from negative effects of CB_offset2 (−0.14) and BSF (−0.07).

Overall, the SHAP analysis clearly reveals that proper band alignment (VB and CB offsets) in both window/absorber and absorber/BSF interfaces, absorber properties, and BSF-related parameters such as electron affinity, bandgap are the primary factors of device performance across all three outputs. [55-56]. The electronic structure and band-alignment engineering mainly dominate PCE and $D^*$ prediction [55] and the optical property such as absorber thickness mainly influence R. This strongly suggests that the precise control of these parameters is essential for achieving optimal device performance.

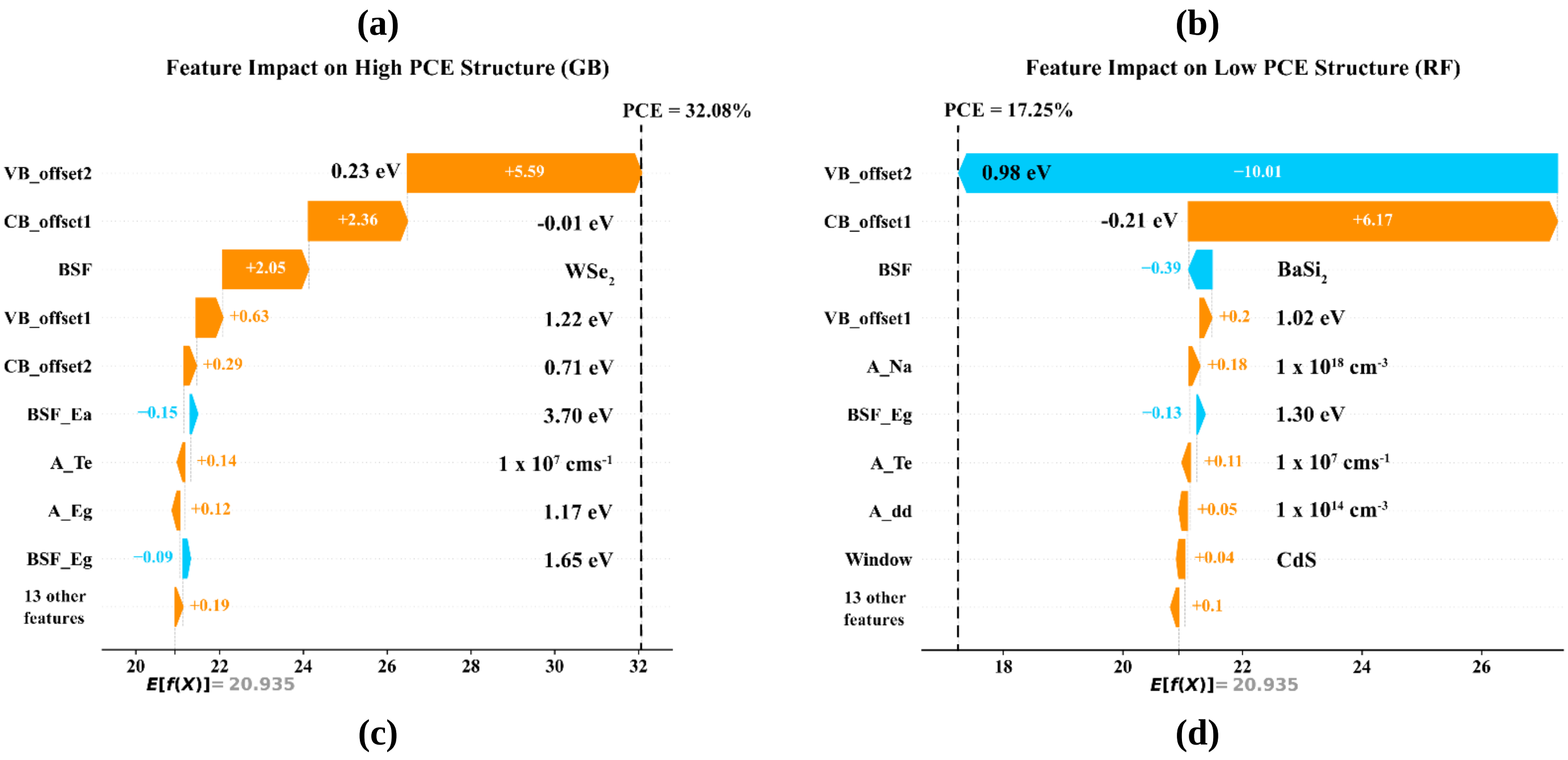

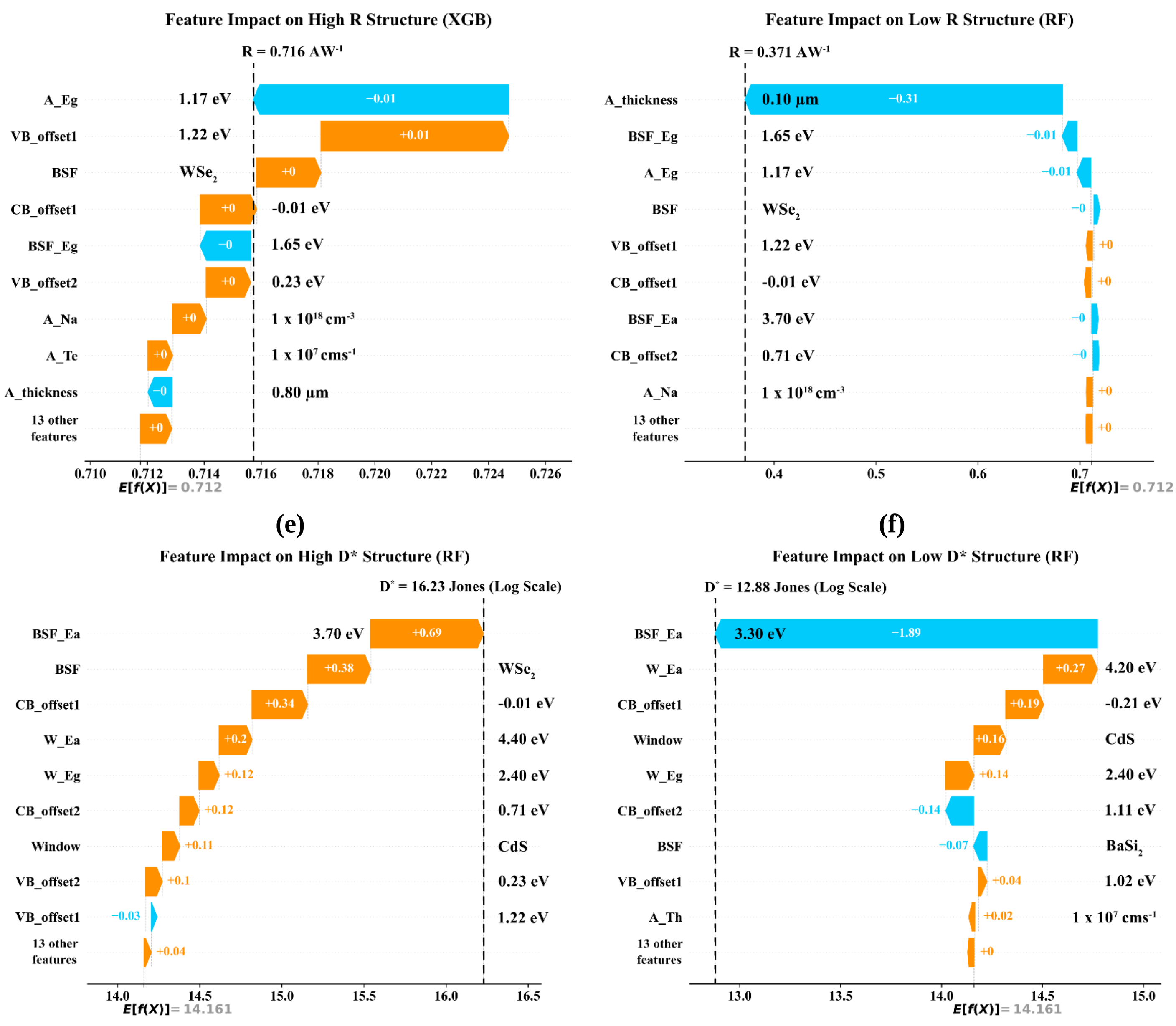

**Figure 6:** SHAP XAI based feature impact analysis showing the contribution of input parameters to high and low performing device structures across the three outputs: PCE (a, b), R (c, d), and $D^*$(e, f).

**(a)** **(b)** **(c)**

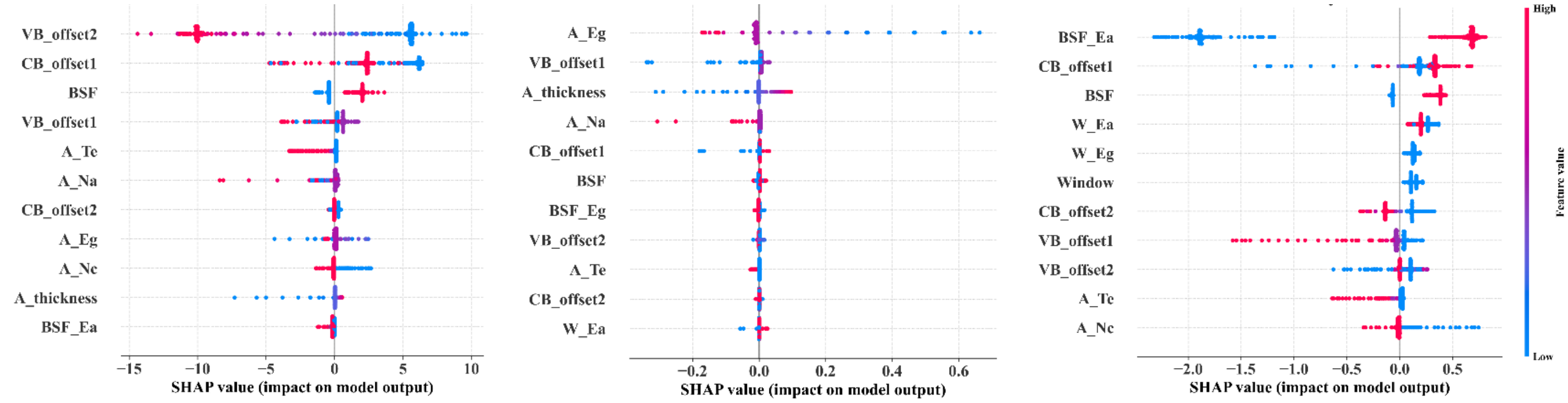

**Figure 7:** SHAP summary plots showing the global feature importance and impact of input parameters on model predictions for (a) PCE, (b) responsivity R, and (c) detectivity $D^*$.

Figure 7 displays the global feature importance and input parameter impacts on model predictions for (a) PCE, (b) responsivity R, and (c) detectivity $D^*$. Each dot represents the SHAP value of a particular feature which indicates its effect to the model output for a single sample. The color bar shows the feature value from low to high. Features are ranked in descending order of importance which highlights the most influential parameters.

For this SHAP analysis, 2 out of the 13 device structures were kept totally unseen by the RF model, representing one high-performing and one low-performing configuration. The RF model accurately predicted the outputs for both cases, showing its strong generalization capability. The SHAP can be utilized beyond global feature impact analysis by introducing parameter level interpretation, through which physically grounded design rules were established [46]. In this study, this approach is further extended to propose and optimize a novel SC-PD device architecture. To provide clearer vision into the SHAP interpretation and its significance to device optimization, Table 7 summarizes the simplified design rules for the SC-PD device. The observed SHAP trends are further converted into practical, physics-guided design rules for efficient device engineering.

**Table 7:** Simplified SHAP derived integrated design rules for the SC-PD device.

| Parameter | Impact on PCE | Impact on Responsivity (R) | Impact on Detectivity ($D^*$) | Integrated Design Rule |
|---|---|---|---|---|
| VB_offset2 | ↑$\Delta E_{V2}$ → PCE↓ Strong negative effect when large VB barrier forms | Minor impact | ↓$\Delta E_{V2}$ → $D^*$↓ Minor impact | Minimize VB_offset2 to improve carrier transport |
| CB_offset1 | Small $\Delta E_{C1}$ → ↑ PCE | Moderate effect | Proper ↑$\Delta E_{C1}$ → $D^*$↑ | Maintain small CB_offset1 for balanced transport and detection |
| $A_N_a$ | ↑$A_N_a$ → PCE↓ | ↑$A_N_a$ → R↓ | Minor effect | Higher absorber doping causes increased recombination losses |
| VB_offset1 | Large $\Delta E_{V1}$ → PCE↓ | ↓$\Delta E_{V1}$ → R↓ | ↑$\Delta E_{V1}$ → $D^*$↓ Strong negative effect when large VB barrier forms | Avoid large VB barriers at absorber/window interfaces |
| $A_E_g$ | Moderate effect | ↓ $A_E_g$ → R↑ | Moderate effect | Use moderate absorber bandgap for improved Responsivity |
| A_thickness | ↓thickness → PCE↓ | ↓thickness → R↓ | Slight effect | Use moderate absorber thickness for efficient absorption |
| $BSF_E_a$ | Small effect | Negligible effect | Optimized ↑$BSF_E_a$ → $D^*$↑ | Optimize BSF electron affinity for higher detectivity |
| A_Te | A_Te ↑ → PCE↓ | Minor effect | A_Te ↑ → $D^*$↓ | Minimize absorber defect density to suppress recombination |
| BSF | BSF ↑→ PCE↑ | Minor impact | BSF↑ → $D^*$↑ | Use optimized BSF layer for efficient carrier collection |

**3.1.4.2 Band Offset Engineering and Optimal Parameter Regimes for SC-PD device**

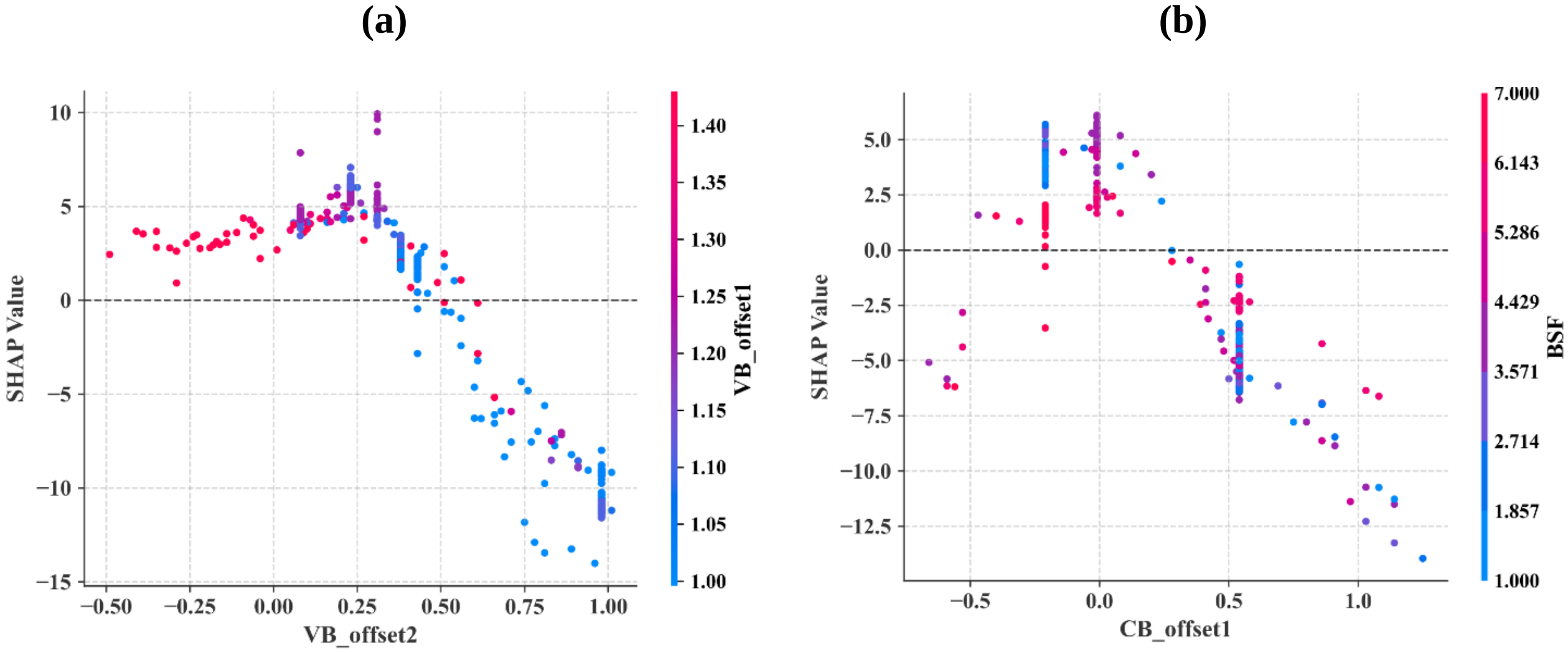

**Figure 8:** SHAP dependence plots explaining the effect of band offset parameters: (a) VB_offset2 and (b) CB_offset1 on the predicted PCE of the SC-PD device.

The SHAP dependence plots for key band-alignment parameters: (a) VB_offset2 and (b) CB_offset1 for the SC-PD device is shown in figure 8. These plots deliver a quantitative interpretation of how variations in individual physical parameters influence the model output through their corresponding SHAP values. Analysis of the positive contribution regions offers valuable insights for optimizing device design strategies.

Figure 8(a) shows a clear non-linear trend for VB_offset2. Here the SHAP values initially increase and reach a positive peak within the range of (~0.1–0.3 eV), which indicates a favorable contribution to device performance. Beyond this region the SHAP value transition from positive to negative. It suggests that higher VB_offset2 values (>0.5 eV) will significantly make a negative impact to the device performance. This behavior highlights the existence of an optimal valence-band offset at the absorber/BSF interface. Similarly, CB_offset1 at figure 8(b) reveals that the parameter shows positive contribution for device efficiency in the range of -0.5 to 0.25 eV, where the peak impact occurs at 0 eV. The SHAP value rapidly declines beyond this range and causes a large negative impact to the device performance. This tendency emphasizes the critical role of conduction-band alignment in achieving efficient carrier selectivity. The color gradients in both plots revels the influence of secondary parameter (VB_offset1 & BSF material) in relation to the primary variables. These interaction revels that the device performance is governed by coupled parameter dependencies rather than individual effects.

These SHAP dependence plots clearly identify the optimal parameter regimes where SHAP values transition from positive to negative, which adds more weight to device design guidelines and performance optimization of SC-PD devices.

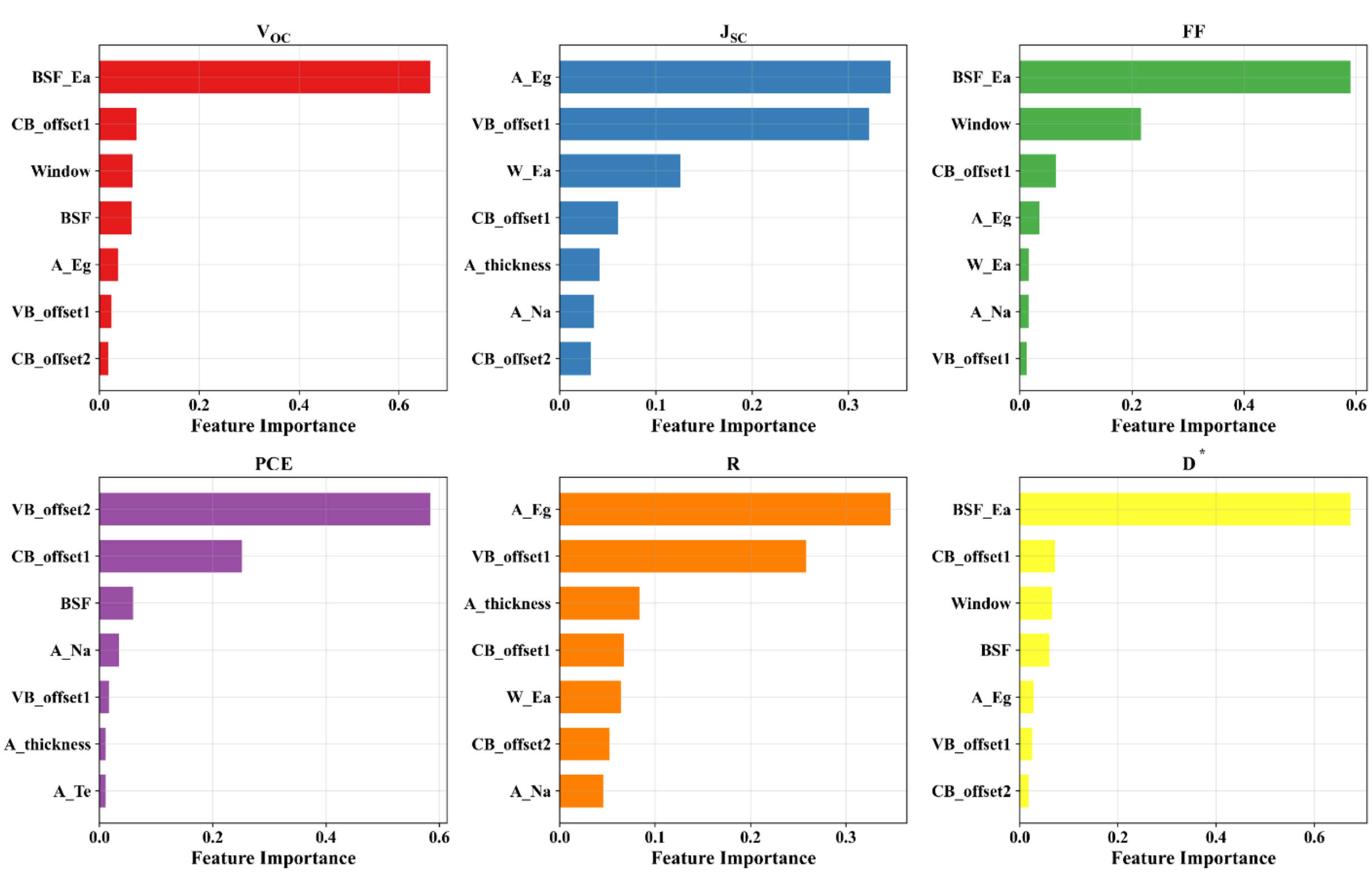

**Figure 9:** The XGB model illustrates Feature importance analysis for six performance parameters.

Figure 9 depicts the feature importance of six individual output parameters ($V_{OC}$, $J_{SC}$, FF, PCE, R, $D^*$) obtained using the XGB model. The figure revels that for predicting $V_{OC}$, BSF_$E_a$ has the largest impact, while the CB_offset1, Window, BSF parameter has negligible influence. For $J_{SC}$, A_$E_g$, VB_offset1 displays the largest contribution, whereas for FF, BSF_$E_a$ has the strongest contribution. The PCE prediction is mostly influenced by VB_offset2 and CB_offset1 which is also seen in the SHAP analysis. On the other hand, A_$E_g$ and VB_offset1 have the most significant effect for predicting R. The detectivity $D^*$ is mainly influenced by the BSF_$E_a$. In general, the XGB-based analysis aligns well with the SHAP-based results, indicating that the most influential physical parameters remain consistent across different models.

### 3.2 $CdSnP_2$ SC-PD: Role of BSF

Figure 10 demonstrates the noteworthiness of the $CuGaSe_2$ (CGS) layer as a BSF layer on $CdSnP_2$ (CTP) SC-PD. Figure 10(a) exhibits the J-V characteristics curve of the CTP SC-PD where in a single heterojunction i.e. n-CdS/p-$CnSnP_2$ the open circuit voltage($V_{OC}$) and short current ($J_{SC}$) are 0.74 V and 33.08 mA/cm$^2$ individually. The implantation of a thin CGS layer with the CTP SC-PD converts the single hetero-junction into a double hetero-junction resulting in the enlargement of performance. The $V_{OC}$ and $J_{SC}$ upsurge to 0.95 V and 39.61 mA/cm$^2$. The credit goes to the BSF layer as it creates a built-in potential with the CTP layer which leads to an increase not only in the $V_{OC}$, but also the $J_{SC}$ by sweeping the generated carriers towards the junction [42].

**(a)** **(b)**

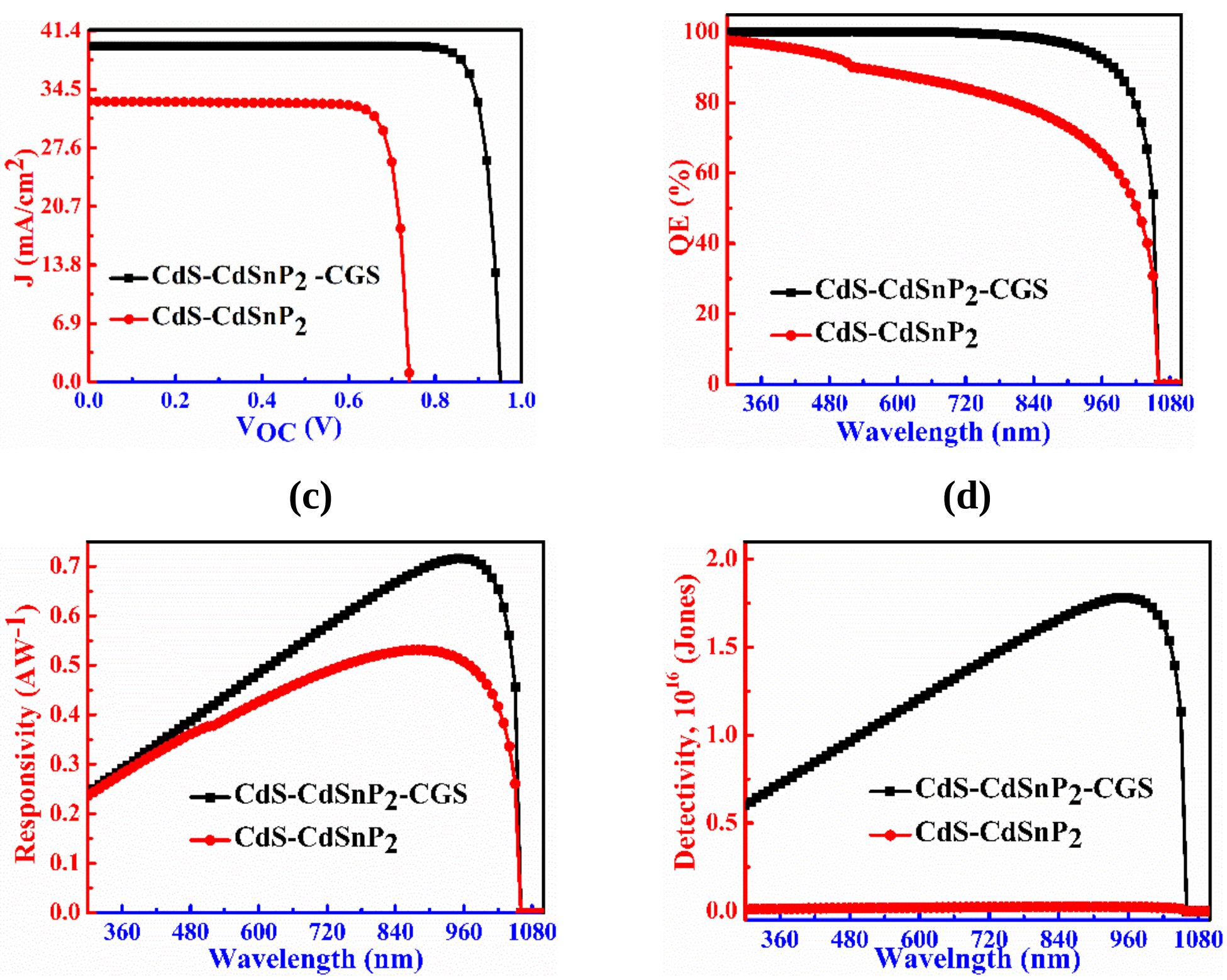

**Figure 10:** $CdSnP_2$ SC-PD (a) J-V, (b) QE, (c) responsivity and (d) detectivity with and without $CuGaSe_2$ BSF panel.

Figure 10(b) exhibits the QE curve of CTP SC-PD. In a single hetero-junction, the photon absorption proficiency of the CTP SC-PD is above 95% and below 98% from wavelength ($\lambda$) 300 to 430 nm, while between 430 and 800 nm, it's about 80-95%. However, after 840 nm, photon absorption capability shrinks linearly and finally just crossing 1060 nm it goes to the 0 value. Interestingly, Photon absorption capability is increased by inserting a thin BSF layer as it holds the maximum value from a shorter wavelength of 380 to a longer wavelength of 1050 nm stimulating the improvement of the Photo-current [40].

The behavior of responsivity in the margin of the wavelength between 300 nm and 1060 nm is depicted in Figure 10(c). In a single hetero-junction, the responsivity of the CTP photodetector obeys a swept advancement with the amelioration of the wavelength and at 880 nm it has reached its peak of .0.53 $AW^{-1}$. Predictably, the appearance of the BSF takes it to the apex of 0.72 $AW^{-1}$ at 950 nm. The amount of the responsivity is measured at the apex point and the credit of its amelioration goes to the progress of the Photon absorption capability of both the shorter and the longer wavelengths [40]. The responsiveness curve depends on the QE Value which can be described by the below equation [8, 57].

$$R = \frac{QE \times e\lambda}{hc} = \frac{J_{SC}}{P_0} \tag{6}$$

Herein, h, c, e, and $P_0$ represent the Planck perpetual ($6.634\times10^{-34}$ J/s), speed of light ($3\times10^{8}$), electron charge ($1.6\times10^{-19}$ C), and occurrence of optical power individually.

Figure 10(d) exhibits the detectivity change with wavelength. Detectivity is measured from the detectivity vs. wavelength curve at the maximum point or peak point. In a Single hetero- junction, CTP PD detectivity is $2.51\times10^{14}$ Jones. By presenting a thin HTL layer, the detectivity ($D^*$) of the CTP PD surges to $1.78\times10^{16}$ Jones. The resulting fact can be described by the below equation (7) [57]:

$$D^* = \frac{R}{\sqrt{(2eJ_0)}} \tag{7}$$

$J_0$ is the reverse or dark current. Equation (7) shows that detectivity is proportional to responsivity. As the R surges so does the $D^*$ also.

### 3.3 Impression of the CTP layer of the $CdSnP_2$ PV-PD

#### 3.3.1 PV parameters by fluctuating depth, impurity and bulk defects of CTP layer

**(a)** **(b)** **(c)**

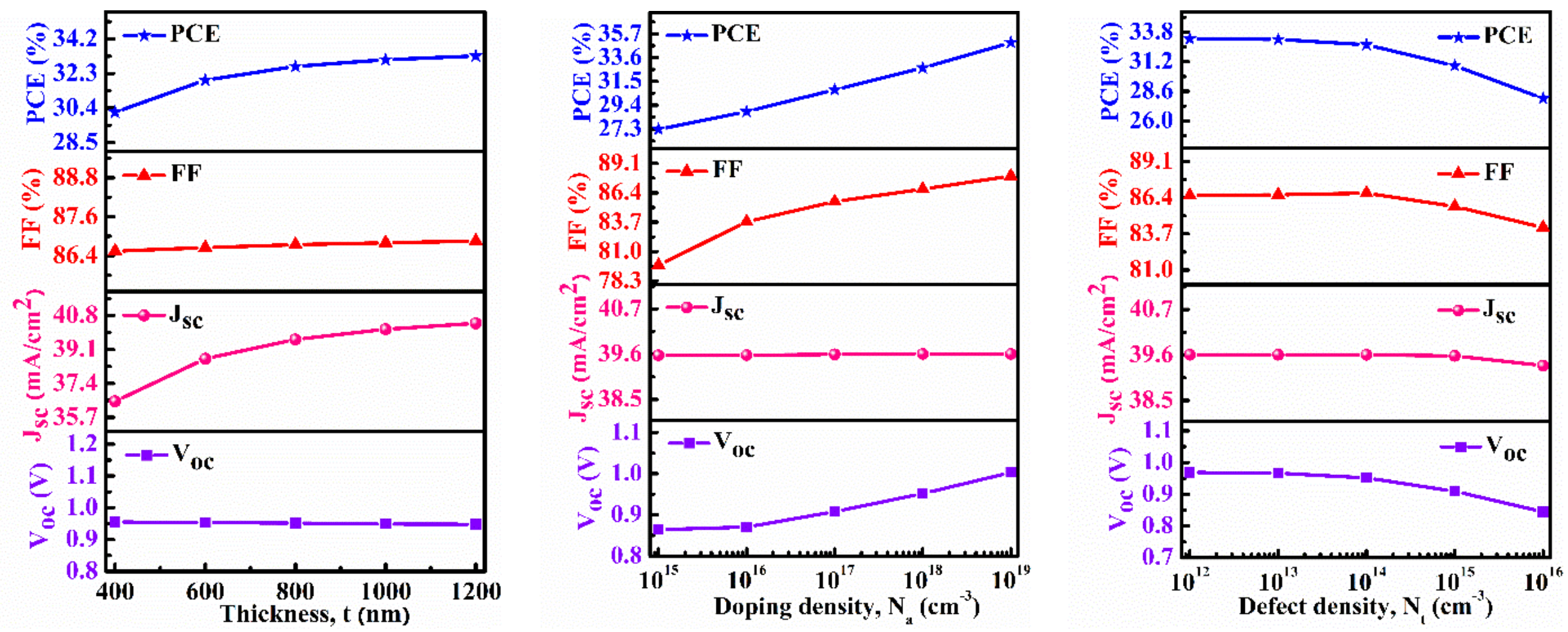

**Figure 11:** PV curve by fluctuating (a) breath (b) Doping and (c) bulk defect , of $CdSnP_2$

The PV characteristics images of the absorber layer of the CTP PV device by fluctuating depth, doping, and bulk defect concentration are displayed in Figure 11. Figure 11(a) demonstrates that by hiking the width up to 1200 nm from 400, $J_{SC}$, FF and PCE rises up significantly from 36.51 to 40.41 mA/cm$^2$, 86.54 % to 86.85% and 30.17 % to 33.26 %, whereas $V_{OC}$ slightly declines from 0.96 to 0.95 V. $J_{SC}$ and PCE upsurges because of a thicker layer enhances the electron-hole pair generation capability [34]. $V_{OC}$ slightly shrinks due to the enlarged thickness of the CTP layer ameliorating recombination as well as dark current which is inversely proportional to $V_{OC}$ [58].

The impression of the doping density upon the CTP layer of CTP PV is revealed in Figure 11 (b). From the figure it is seen that when the level of dopant concentration hikes from a lower concentration of $10^{16}$cm$^{-3}$ to a higher concentration of $10^{20}$ cm$^{-3}$, the performance parameter $J_{SC}$ increase a little from 39.57 to 39.60 mA/cm$^2$. On the reverse side, there is an annexation of the $V_{OC}$, FF and PCE from 0.87 to 1.06 V, 83.74 % to 88.6% and 28.84 % to 37.22 %. The recombination current decreases with an increase in acceptor doping due to the elevation of built-in potential, thereby enhancing the $V_{OC}$ [40].

Figure 11(c) displays the impact of the defect on the CTP layer of the CTP PV. The defect disparity from $10^{12}$ to $10^{14}$ $cm^{-3}$ doesn't effect on the $J_{SC}$ of the PV device. The values $J_{SC}$ are correspondingly 39.61 mA/$cm^2$. When the bulk defect increases from$10^{15}$ to $10^{16}$ $cm^{-3}$ , $J_{SC}$ slightly declines from 39.58 to 39.34 mA/$cm^2$ which is not so worthy to consider. The value of $V_{OC}$, FF and PCE decreases from 0.97 to 0.84 V, 86.57 % to 84.17% and 33.21 % to 27.97 % with increasing defect from $10^{12}$ to $10^{16}$ $cm^{-3}$. However, beyond our observed range, the performance of the CTP PV may decline with increasing bulk defects as it enlarges the Shockley-Read-Hall (SRH) recombination which is not benign for the performance parameter [40].

### 3.3.2 R, and $D^*$ by fluctuating depth, impurity and bulk defects of CTP layer

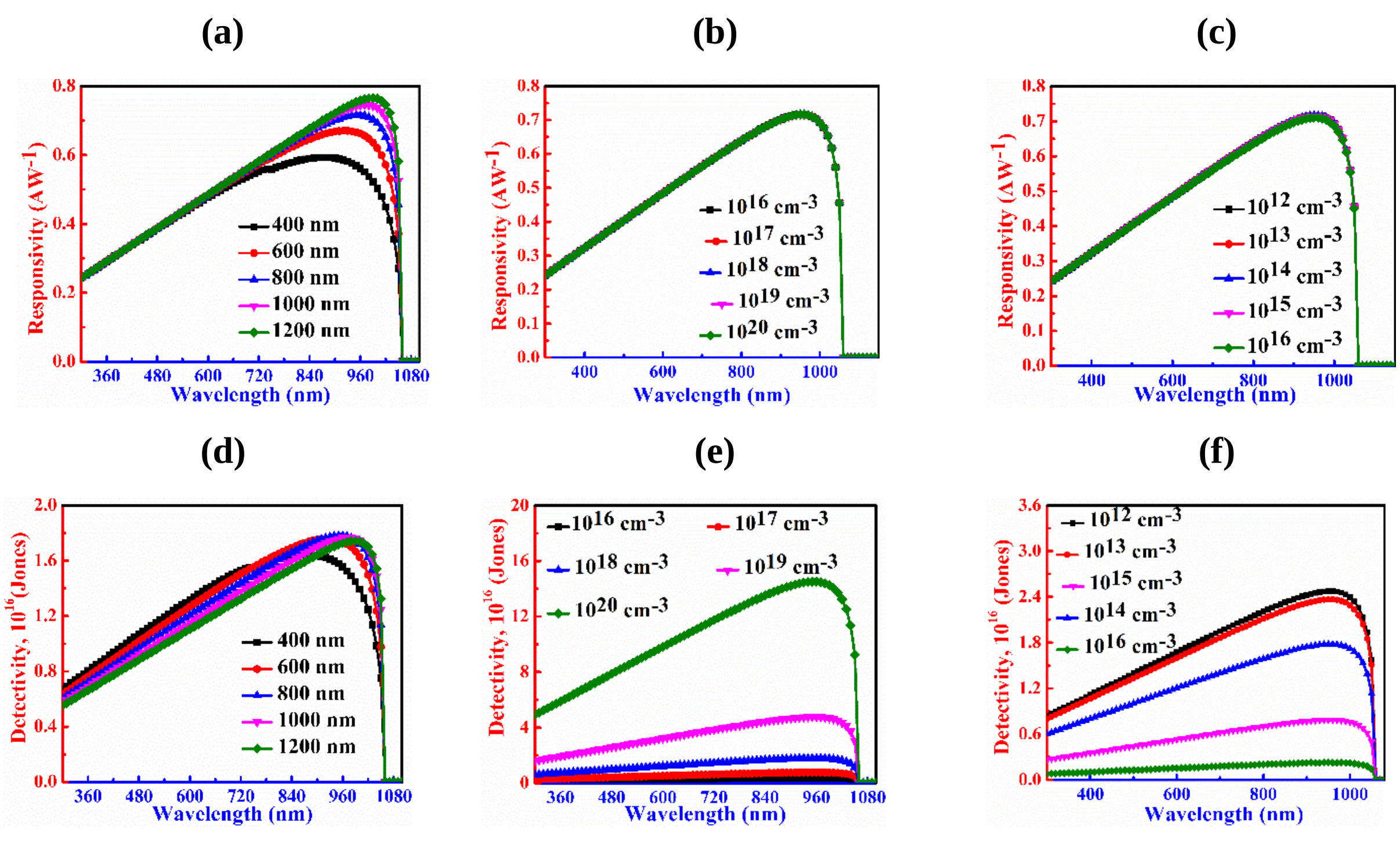

**Figure 12:** R by fluctuating (a) thickness (b) doping concentration (c) defect density, $D^*$ by fluctuating (d) thickness (e) doping concentration and (f) defect density, of $CdSnP_2$.

The R, and $D^*$ characteristics images of the absorber layer of the CTP PD by fluctuating depth, doping, and bulk defect concentration are displayed in Figure 12. Figure 12(a) shows how different thicknesses of the CTP layer affect the performance of the CTP PD. Thickness varies from 400 to 1200 nm with a span of 200 nm difference. Surprisingly, the impact of the thickness is observed in the wavelength between 880 and 990 nm. When the breadth hikes from 0.400 to 1.200 μm of the CTP layer, R hikes in between 0.59 to 0.77 A/W.As the absorber layer thickness, the responsiveness increases due to longer wavelength electron-hole absorption and higher generation rates. This is also accompanied by longer lifetimes and diffusion lengths [40, 59]. Stated differently, a thicker absorber layer generates greater current because it absorbs more photons than a narrower layer. As a result, the responsivity increases since it is related to the increase in photocurrent [57].

Figure 12(b) gained by fluctuating doping concentrations of the CTP layer. From the figure, one can realize that doping concentrations of CTP layer doesn't impact on the responsivity of CTP PD. The value of responsivity is 0.72 $AW^{-1}$ at 950 nm.

Figure 12(c) demonstrates that bulk defect concentration distinction from $10^{12}$-$10^{16}$ $cm^{-3}$ of the CTP layer does not affect the responsivity performance of the CTP photodetector. The PD performance parameters R is 0.72 A/W .

In Figure 12(d), Detectivity altered with wavelength by fluctuating thickness from 400 to 1200 nm of the CTP layer. When the width rise in between 0.400 to 1.200 μm, $D^*$ lessened slightly from (1.64-1.74) × $10^{16}$ to 1.74×$10^{16}$ at a wavelength range upto 990 nm from 880 nm. Detectivity declines with increased thickness because of higher thickness increased amount of dark current which can be described by the following equation (8) [57,60].

$$J_0 = \frac{Jsc}{e^{\frac{Voc}{Vt}} - 1} \quad (8)$$

The Value of thermal voltage ($V_t$) is 26 mV. As the dark current rises, it declines the detectivity because detectivity is in reverse relational to the root of the dark current ($J_0$) from equation (7).

Figure 12(e) illustrates that detectivity changes with wavelength by fluctuating the doping concentration of the CTP photodetector. At minor doping density $D^*$ is small and at upper doping $D^*$ is huge. This generally happens because at higher doping $J_0$ is low [61]. From equation (7) the detectivity is inversely proportional to dark current as a result detectivity is higher at higher doping. When the doping hiked from $10^{16}$ to $10^{20}$ cm$^{-3}$ detectivity increased from $3.72\times10^{15}$ to $1.45\times10^{17}$ Jones [57].

Figure 12 (f) illustrates detectivity change with wavelength by fluctuating defect concentration of the CTP layer of the CTP PD. Detectivity decreases from $2.47\times10^{16}$ to $2.27\times10^{15}$ Jones with the increasing defect from $10^{12}$ to $10^{16}$cm$^{-3}$ . This is because at upper bulk defects SHR recombination is greater as a result $J_{SC}$ decreases but $J_0$ escalations and at that time R is constant for all defects [36]. Equation (7) shows that detectivity is contrary to the root of the $J_0$ as a result detectivity declines with growing defects [62].

### 3.4 Performance of $CdSnP_2$ Photodetector with CdS window layer

#### 3.4.1 PV parameters by fluctuating depth, impurity and bulk defects of CdS layer

**(a)** **(b)** **(c)**

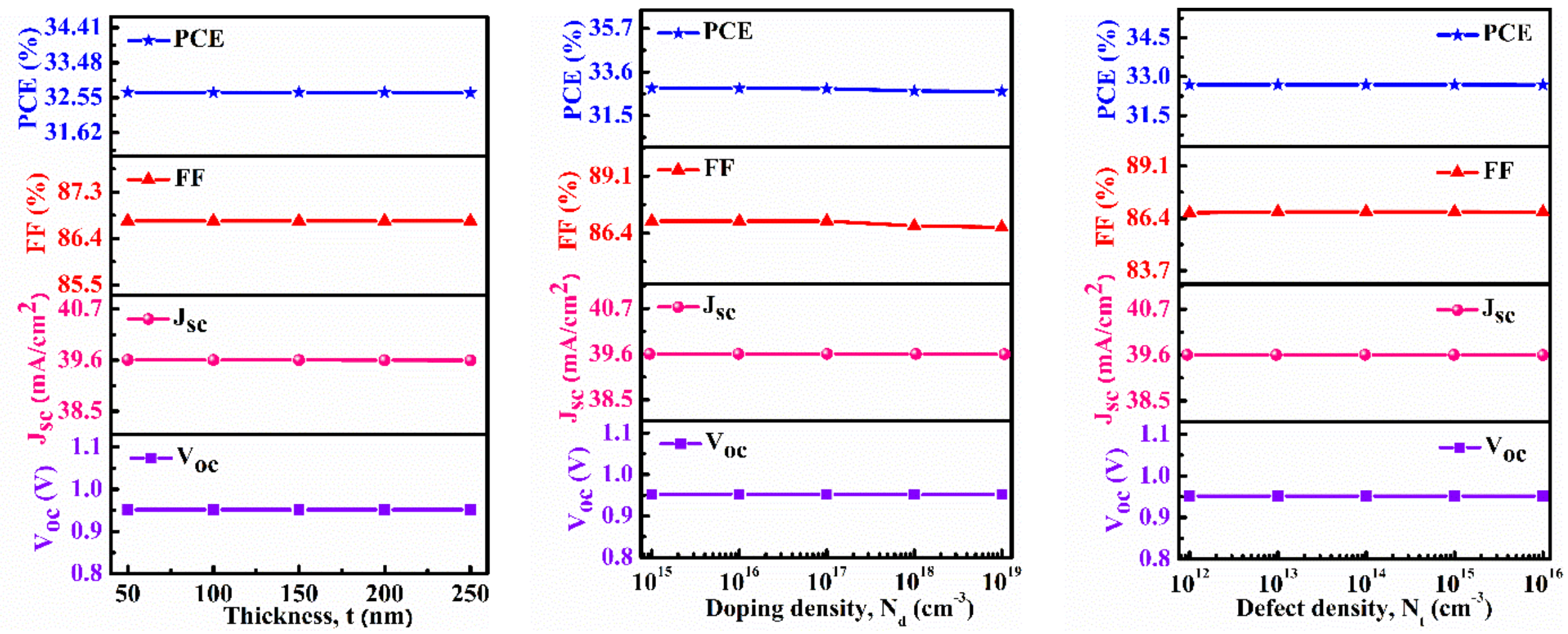

**Figure 13:** PV characteristics curve by fluctuating (a) Breadth (b) impurity density (c) bulk defect, CdS of the ETL of CTP Photodetector.

The window layer plays a vital role in the CTP PV cells. Figure 13 exhibits the impression of distinction depth, impurity, and bulk defect concentration of CdS on the CTP SC. In Figure 13(a), $V_{OC}$, $J_{SC}$, FF and PCE appear to be constant when the depth of CdS varies from 50 nm to 250 nm i.e. depth variation of CdS does not affect CTP SC noticeably. The values of PV metrics are correspondingly 0.95 V, 39.61 mA/cm$^2$, 86.74 % and 32.69 % respectively.

Discrepancies in the doping concentration of CdS slightly impact of performance of the CTP SC which is presented in Figure 13(b). When the doping varies from $10^{15}$ to $10^{18}$ cm$^{-3}$, PV parameters appear to be constant and the values of PV parameters are correspondingly 0.95 V, 39.61 mA/cm$^2$ , 86.97 % and 32.81 %. However, when doping increased to $10^{19}$ cm$^{-3}$, $J_{SC}$, FF and PCE a touch decreased to 39.60 mA/cm$^2$, 86.68 % and 32.66 %, whereas $V_{OC}$ remained as same as before. This is because at higher doping recombination is higher which declines appliance performance [40].

Designing any film or PV cell defect is an important factor. Figure, 13(c) displays the effect of defect density of CdS on the CTP SC. Bulk defect concentration from $10^{12}$ to $10^{16}$ cm$^{-3}$ does not affect the CTP SC. The values of PV parameters are correspondingly 0.95 V, 39.61 mA/cm$^2$, 86.74 % and 32.69 % respectively. .

### 3.4.2 R and $D^*$ by fluctuating depth, impurity and bulk defects of CdS layer

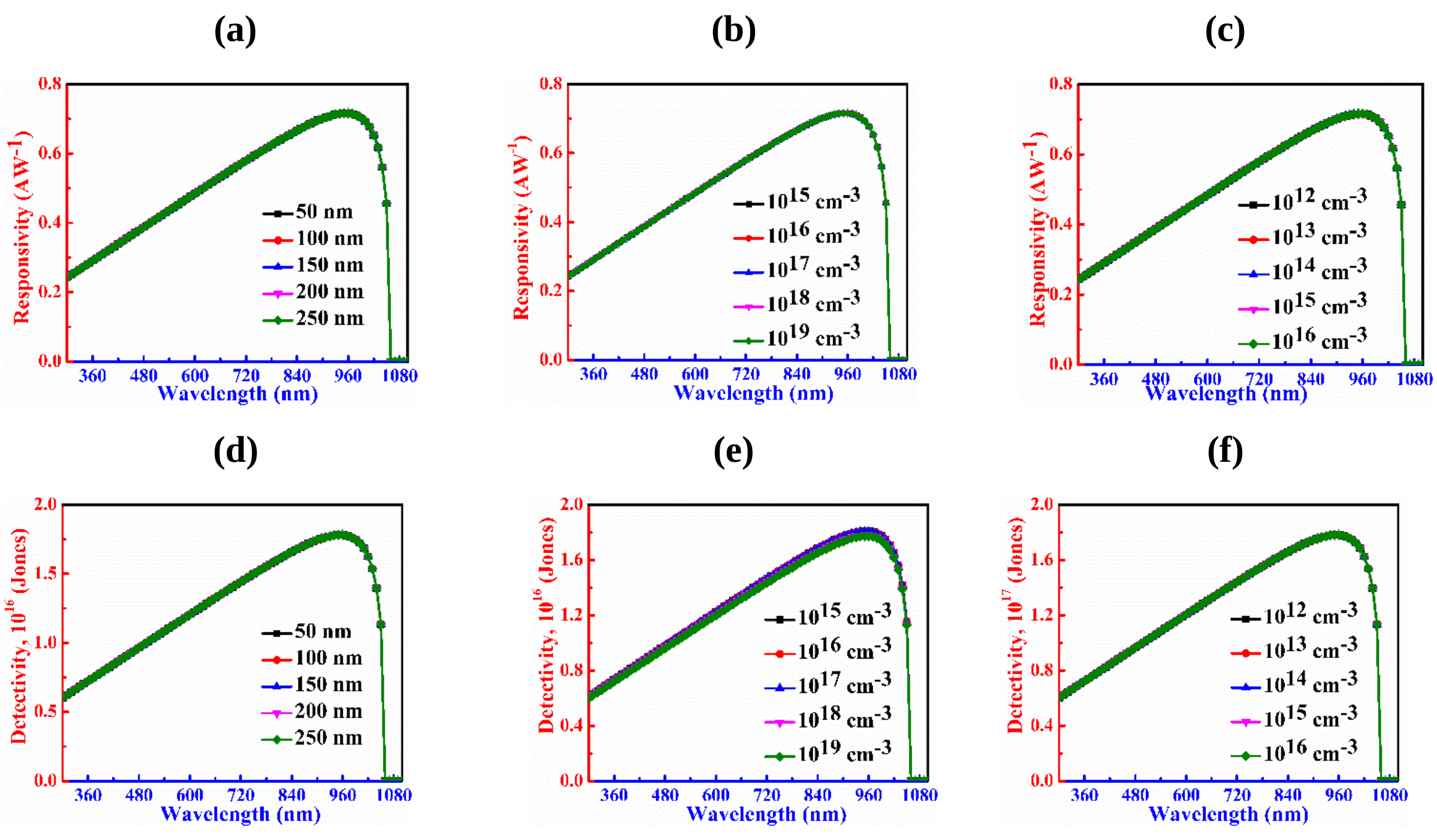

**Figure 14:** R by fluctuating (d) thickness (e) doping concentration (f) defect density, $D^*$ by fluctuating (g) thickness (h) impurity concentration (i) Bulk defect n, CdS of the window of CTP Photodetector.

Figures 14(a-f) display that the disparity of thickness, doping concentration, and defect density of CdS doesn't affect the responsivity and detectivity performance of the CTP PD. The values of responsivity and detectivity are 0.72 A/W and 1.78×10$^{16}$ Jones at 950 nm wavelength.

## 3.5 Performance CGS as a hole transporting layer (ETL) layer on $CdSnP_2$ PD

### 3.5.1 PV parameters by fluctuating depth, impurity and bulk defects of CGS level

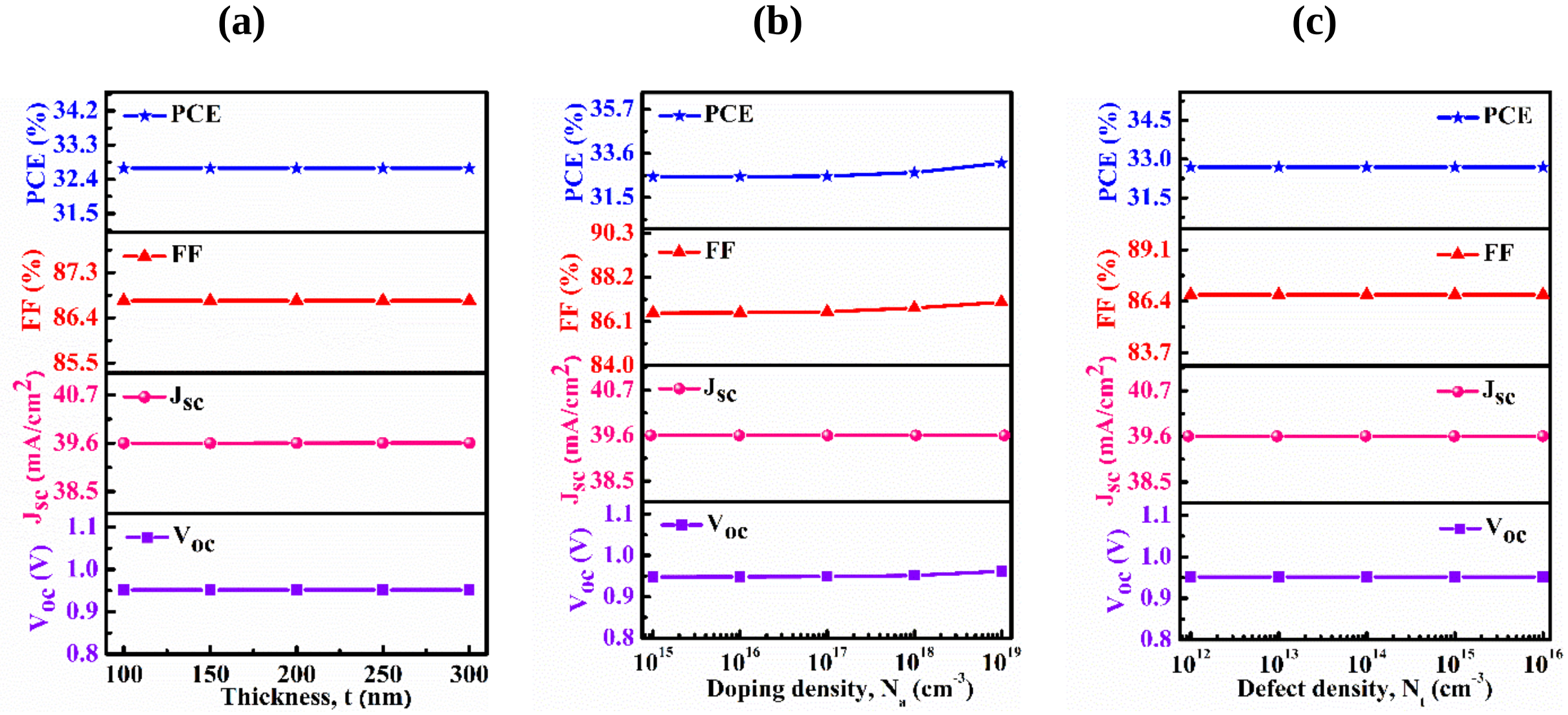

**Figure 15**: PV characteristics curve by fluctuating (a) Width (b) impurity concentration (c) bulk defect density, CGS of the BSF level of CTP PV cell.

Figure 15 shows the influence of the variation of thickness, doping, and defect of the CGS layer or BSF layer on the CTP SC. Figure 15(a and c) shows that variation in depth and bulk defect concentration does not impact of performance of the CTP PV. Thickness varies from 100 to 300 nm, and bulk defect concentration varies from $10^{12}$ to $10^{16}$ cm$^{-3}$. The values of the PV parameters are correspondingly 0.95 V, 39.61 mA/cm$^2$, 86.74 % and 32.69 %. On the other hand, doping concentration increase from $10^{15}$ to $10^{19}$ cm$^{-3}$, $V_{OC}$, FF and PCE increase from 0.95 V to 0.96 V, 86.50 % to 87.01 % and 32.48 % to 33.14 % whereas $J_{SC}$ seems to be almost constant. Higher doping create a strong build in potential which increase the $V_{OC}$. As $V_{OC}$ surges while $J_{SC}$ remains nearly constant, both FF and PCE also increase. This occurs because FF is depend on $V_{OC}$, while PCE depends on $J_{SC}$, $V_{OC}$, and FF [42].

### 3.5.2 R and $D^*$ by fluctuating depth, impurity and bulk defects of CGS layer

(a) (b) (c)

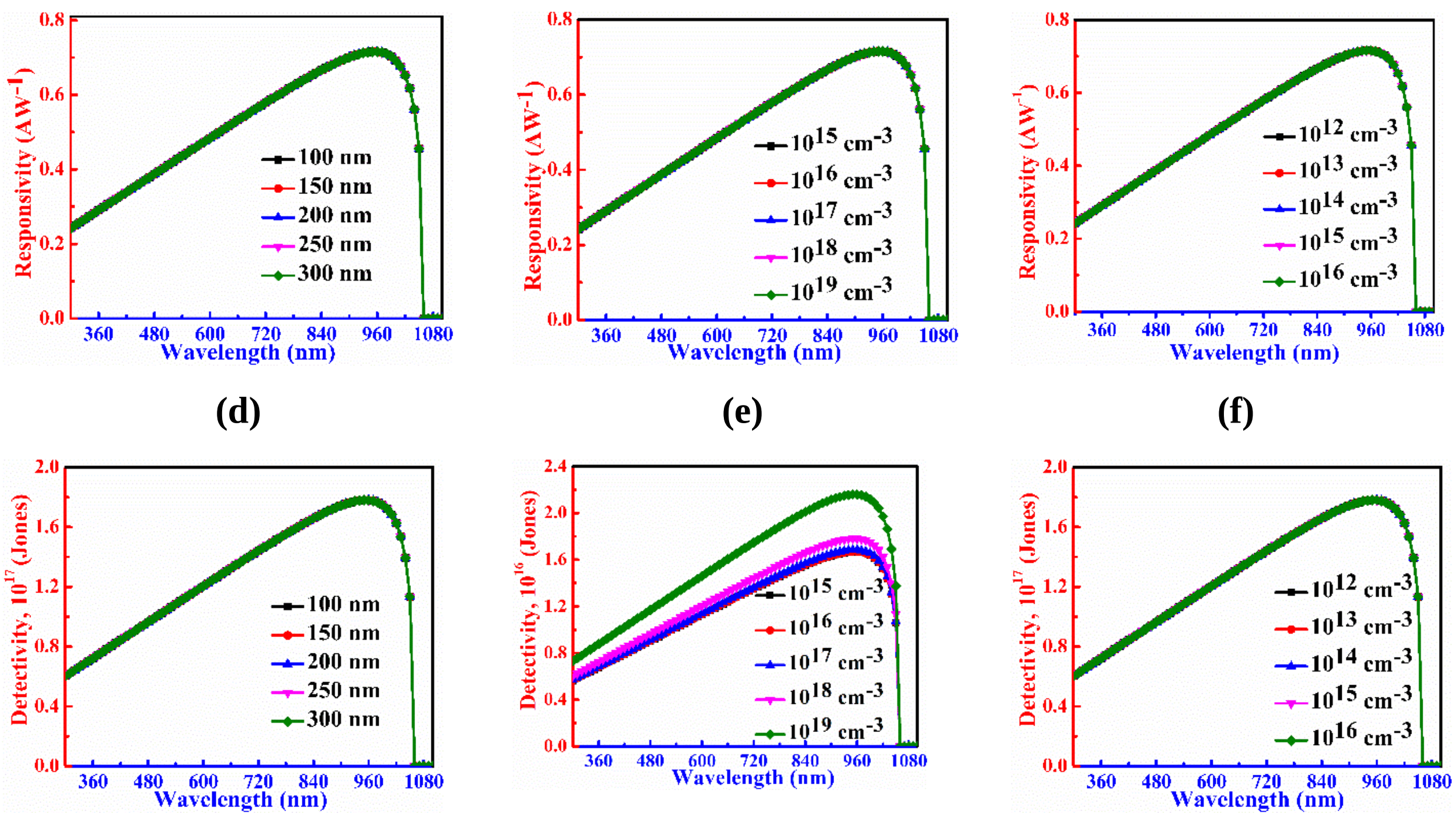

**Figure 16**: R curve by fluctuating (a) thickness (b) doping concentration (c) defect density, $D^*$ by fluctuating (d) thickness (e) impurity concentration (f) bulk defect density, CCG of the BSF layer of CTP Photodetector.

Figure 16 shows the influence of the variation of thickness, doping, and defect of the CGS layer or BSF layer on the CTP PD. Figure 16( a -f) shows that variation in depth, doping and bulk defect concentration does not impact of performance of the CTP PD except doping concentration. Thickness varies from 100 to 300 nm, and bulk defect concentration varies from $10^{12}$ to $10^{16}$ $cm^{-3}$. The value of the R is 0.72 A/W at 950 nm. The value of the detectivity for variation of thickness and defect density is 1.78 $\times 10^{16}$ Jones at wavelength 950 nm. However, the acceptor concentration of the BSF affects the detectivity of the CTP PD. At lower doping i.e. $10^{15}$ to $10^{17}$ $cm^{-3}$, the detectivity of the CTP layer is slightly increased to (1.67-1.68) $\times 10^{16}$ Jones at 950 nm wavelength. At higher doping between $10^{18}$ and $10^{19}$ $cm^{-3}$ detectivity increases from $1.78\times10^{16}$ to $2.16\times10^{16}$ Jones at 950 nm. Dark current decreases with increasing doping concentration, which increases detectivity [63-64]. By considering many properties optimum thickness, doping concentration, and defect density correspondingly 200 nm, $10^{18}$ $cm^{-3}$, and $10^{14}$ $cm^{-3}$.

### 3.6 Temperature effect on $CdSnP_2$ Photodetector

The CTP SC-PD can operate in different temperatures; it is essential to determine performance parameters at different temperatures. Figure 17(a) shows that when temperature varies from 250 to 350 K, $J_{SC}$ keeps on almost constant at 39.61 $mA/cm^2$ but $V_{OC}$, FF and PCE slightly decreases from 0.97 to 0.92 V, 88.56 % to 84.8% and 33.86 % to 30.83% at higher temperatures. The current shows a static behavior may be due to a higher temperature; upsurges non-radiative recombination (like Shockley-Read-Hall) can offset the thermally produced carriers. Furthermore, if internal resistances increase with temperature, they can bound charge collection, preventing the expected rise in $J_{SC}$. The reason behind the decrement of the $V_{OC}$ is the betterment of the dark current at higher temperatures. As $V_{OC}$ declines while $J_{SC}$ remains nearly constant, both FF and PCE also decrease. This occurs because FF is directly depended on $V_{OC}$, while PCE depends on $J_{SC}$, $V_{OC}$, and FF [40].

**(a)** **(b)**

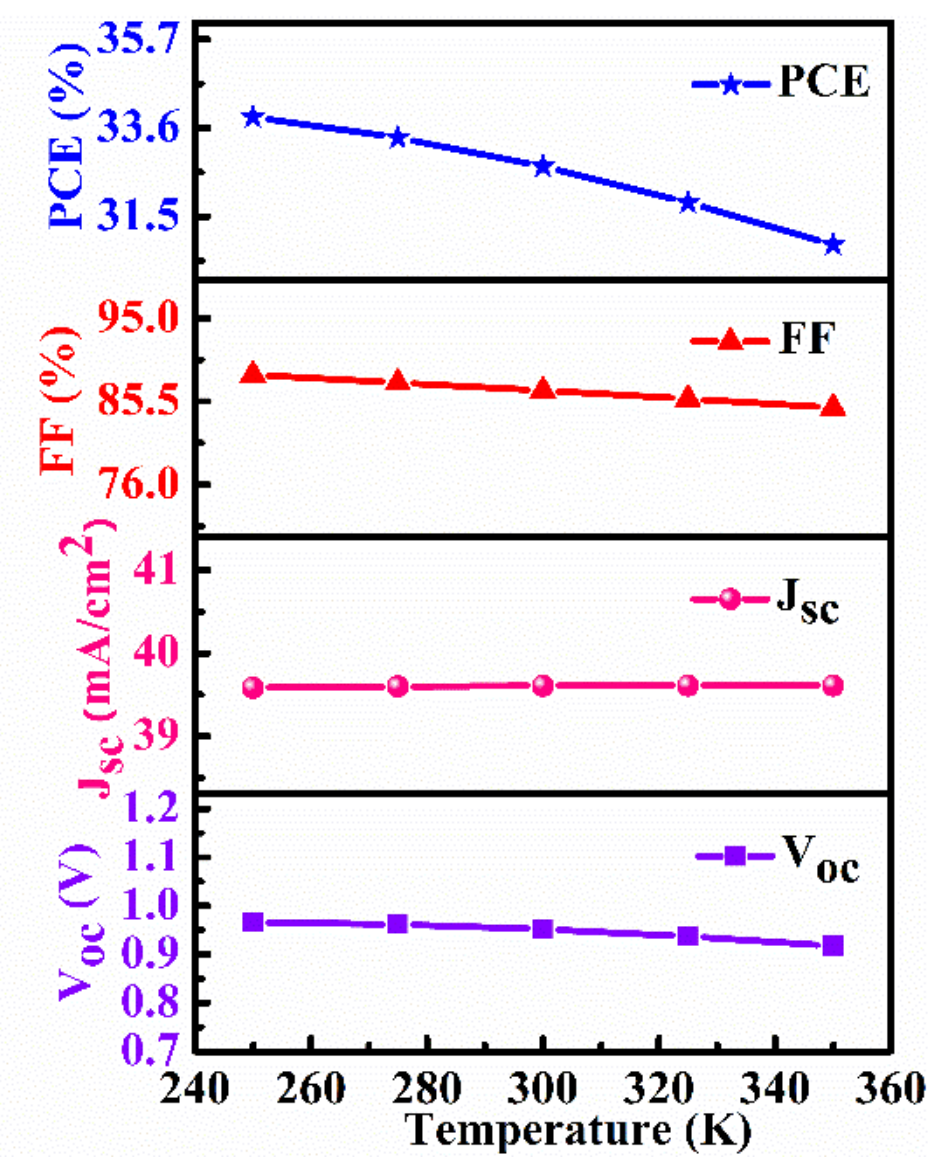

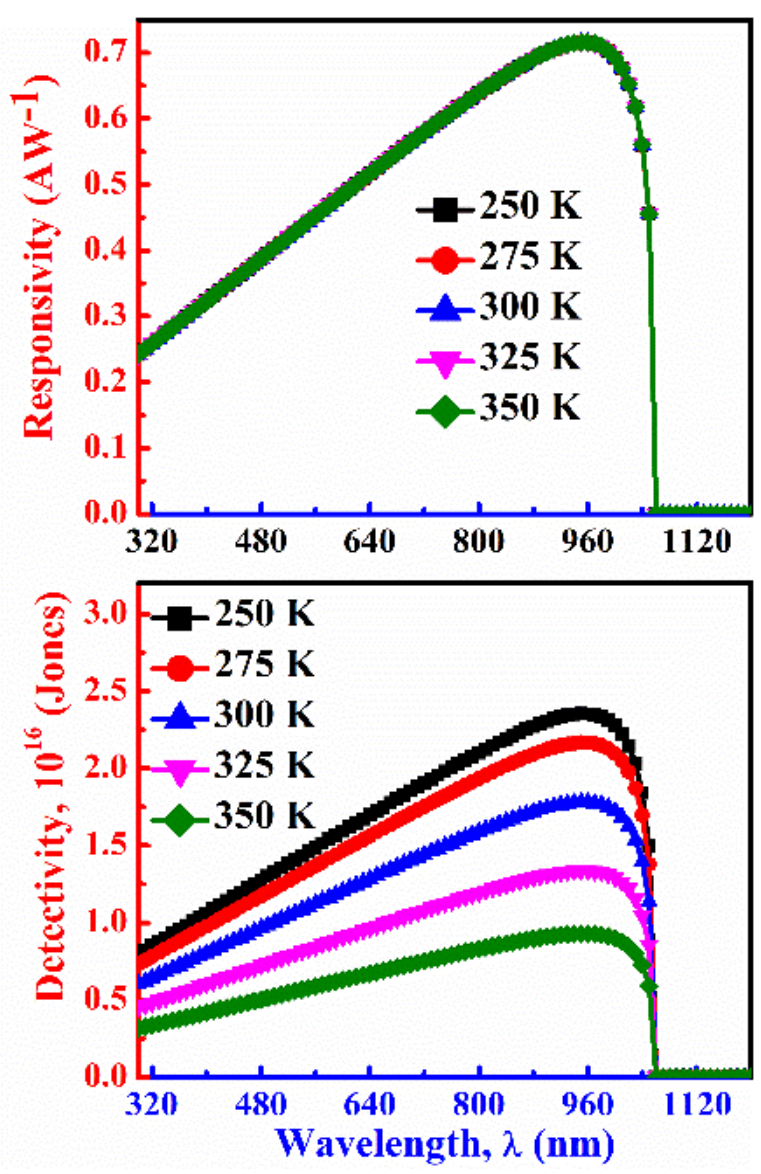

**Figure 17:** (a) PV and (b) PD characteristic curve of n-CdS /p-$CdSnP_2$/ p+-$CuGeSe_2$ PD by fluctuating temperature.

Figure 17(b) displays that the photodetector performance parameters with the increase of temperature. Responsivity shows a constant behavior with the surge of temperature whereas $D^*$ decrease from (2.35-9.30)$\times 10^{16}$ Jones with rise of temperature from 250 K to 350 K. At higher temperature dark current is higher which decrease the detectivity [40].

### 3.7 Resistance effect on $CdSnP_2$ Device

#### 3.7.1 Series and shunt resistance impact on CTP PV device

**(a)** **(b)**

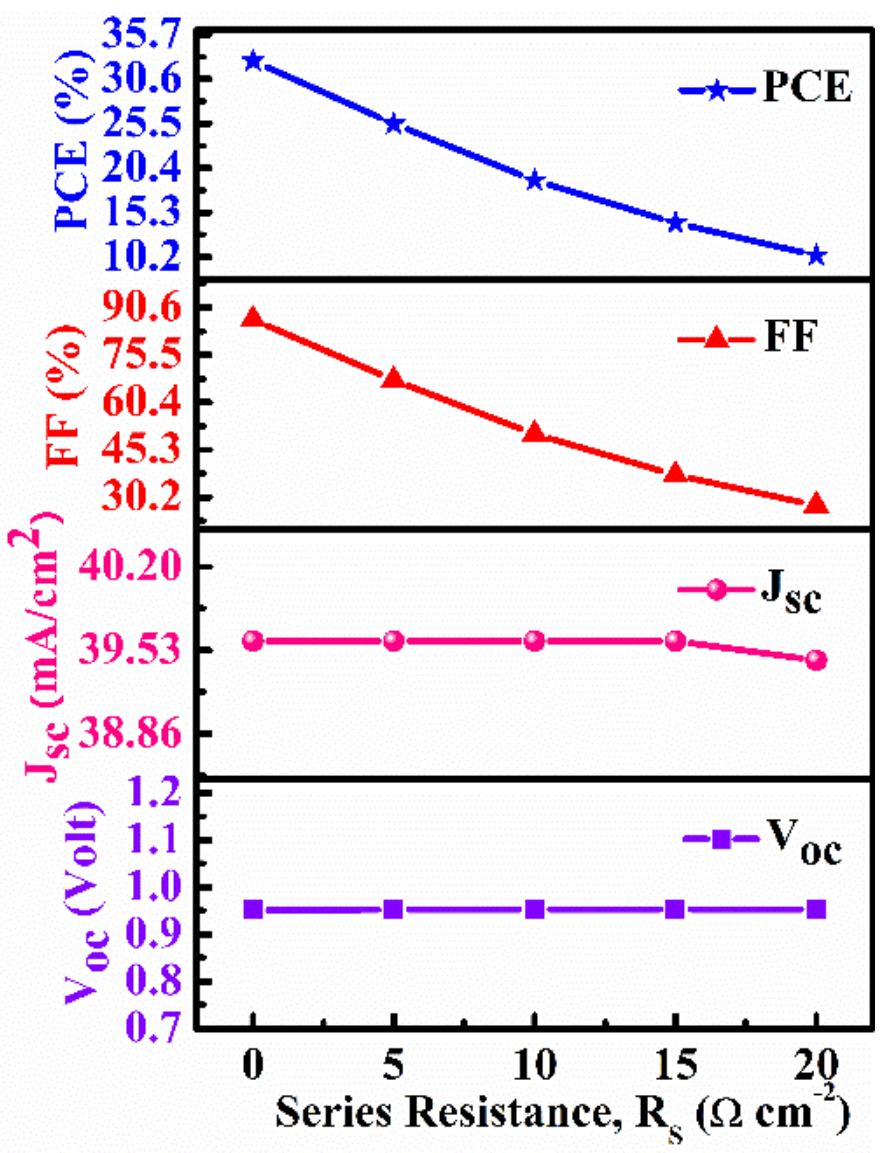

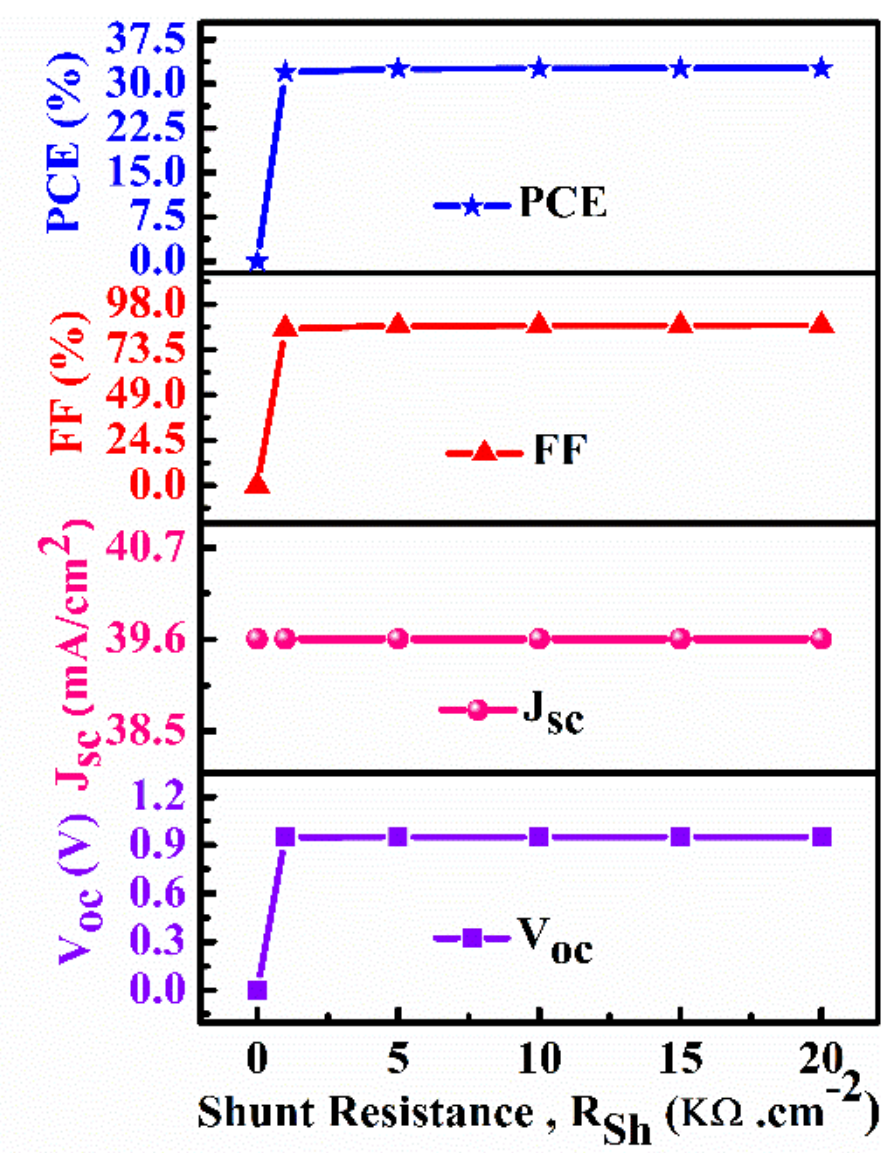

**Figure 18:** PV characteristic curve of n-CdS /p-$CdSnP_2$/ p+-$CuGeSe_2$ Photodetector by fluctuating (a) Series and (b) Parallel Resistances.

Figure 18 illustrates the impact of series and parallel resistance on the performance of CTP SC. The series resistance is examined within a range of 0 Ω.cm² to 20 Ω.cm², where $V_{OC}$ remains nearly constant at 0.95 V. The $J_{SC}$ experiences a slight decrease from 39.61 mA/cm² to 39.44 mA/cm², while FF and PCE decline significantly from 86.74 % to 27.66 % and 32.69 % to 10.4 %, respectively. Series resistance main impact is a reduction in the FF, though excessively high resistance can also lower the short-circuit current. Additionally, a high series resistance can lead to heating effects, reducing the overall efficiency and stability of the device [45].

Shunt resistance ($R_{SH}$) in solar cells, often caused by manufacturing defects, leads to considerable power losses. A low shunt resistance provides an alternate path for the light-generated current, reducing the current flow through the junction and lowering the cell's voltage. The influence of parallel resistance, fluctuating within a range of 0 to 20 kΩ.cm², is also observed. As parallel resistance increases from 0 to 1 kΩ.cm², the $V_{OC}$, FF, and PCE improve to 0.95 V, 84.88 %, and 31.96 %, respectively, while $J_{SC}$ remains unchanged at 39.61 mA/cm². Further increases in resistance, from 5 to 20 kΩ.cm², result in the PV parameters maintaining an almost constant level. This stability in PV parameters suggests that beyond a certain threshold, the effect of parallel resistance does not contribute significantly to performance enhancement [45].

### 3.7.2 Series and shunt resistance impact on CTP PD device

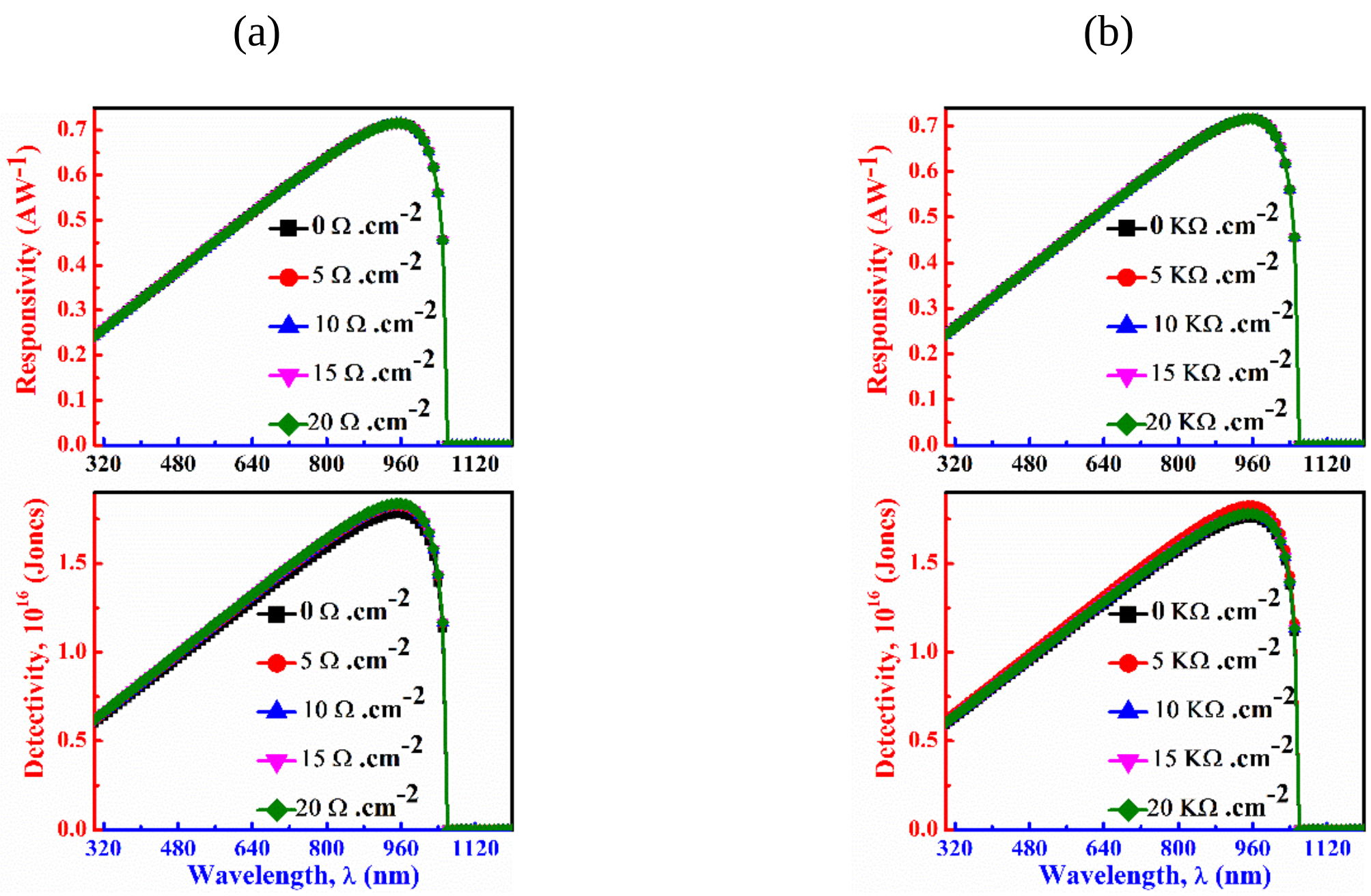

**Figure 19:** PD characteristic curve of n-CdS /p-$CdSnP_2$/ p+-$CuGeSe_2$ Photodetector by fluctuating (a) Series and (b) Parallel Resistances.

The CTP PD R and D* curve is shown in Figure 19. The figure shows that while detectivity has little influence, R appears to be nearly constant as series and shunt resistance grow.

### 3.8 Overall Competence

Table 8 presents the gadget's performance both with and without a thin CGS layer as the BSF layer, with CdS serving as the electron-transporting or window layer. The inclusion of the CGS layer significantly enhances the PV and PD performance parameters.

**Table 8:** Displayed overall performance of the CTP photodetector with and without the CGS layer.

| **Appliance Structure** | **PCE (%)** | **Responsivity (A/W)** | **Detectivity (Jones)** | **Wavelength (nm)** |
|---|---|---|---|---|
| n-CdS/p-$CdSnP_2$ | 20.67 | 0.53 | $2.51\times10^{14}$ | 880 |
| n-CdS/p-$CdSnP_2$/p+-CGS | 32.68 | 0.72 | $1.78\times10^{16}$ | 950 |

Table 9 presents a comparative investigation of various experimental and theoretical hetero-junction SC-PD cell structures that exploit different absorber layers. The data indicates that SC-PD cells with fluctuating absorber layers exhibit a PCE range of 3.87-29.1%, a responsivity range of 0.017 to 0.6345 A/W, and a detectivity range of $3.9 \times 10^9$ to $4.08 \times 10^{13}$ Jones. In contrast, our proposed solar cell-photodetector, which incorporates a CTP-based absorber layer, achieves a PCE 32.68 %, responsivity of 0.72 A/W and an exceptionally high detectivity of $1.78 \times 10^{16}$ Jones. Based on these values, the proposed structure presents a significant advancement and contributes to the diversity of efficient designs in the scientific record.

**Table 9:** Comparison between experimental and simulated photodetector gadgets.

| Structure | Efficiency (%) | Responsivity ($AW^{-1}$) | Detectivity (Jones) | Type | Reference |
|---|---|---|---|---|---|
| CdS/$CdSnP_2$/CGS | 32.68 | 0.72 | $1.78\times10^{16}$ | Simulation | This report |
| $WS_2/AlO_x$/Ge | - | 0.6345 | $4.3 \times 10^{11}$ | Experimental | 19 |
| Si | 10.2-27.1 | 0.52 | $3.26 \times 10^{13}$ | Experimental | 20, 6 |
| GaAs | 18.4-29.1 | 0.262 | $1 \times 10^{12}$ | Experimental | 22, 6 |

| Graphene/Si | - | 0.73 | $5.77 \times 10^{13}$ | Experimental | 24 |
|---|---|---|---|---|---|
| Mg-$ZnSnP_2$-Sn | 3.87 | 0.2 | $4.62\times 10^{12}$ | Experimental | 25, 65 |
| PbS/$TiS_3$ | - | 0.36 | $3.9 \times 10^{9}$ | Simulation | 26 |
| CdTe | 21 | - | - | Experimental | 6 |
| OPV | 14.5-15.8 | | | Exp./Sim | 6 |

## 4. Conclusion

A novel CTP-based solar cell–photodetector (SC–PD) has been proposed and optimized using a hybrid SCAPS–ML framework by systematically evaluating thirteen different CTP-based device configurations. Among these architectures, the CdS and CGS layers were identified as the optimal window and BSF layers, respectively, in agreement with the predictions of the ML-assisted optimization framework. Furthermore, SHAP analysis revealed that band-offset engineering, particularly the valence-band offset at the absorber/BSF interface VB_offset2 and CB_offset1and the conduction-band offset at the window/absorber interface, CB_offset1, together with the BSF activation energy, BSF_$E_a$ and absorber properties (A_$E_g$ and A_thickness), are the dominant factors governing device performance. The SHAP interpretation also enabled the establishment of physics-guided design rules for device optimization. After optimizing the physical parameters of the constituent layers, the incorporation of a thin 200 nm CGS BSF layer significantly enhanced both photovoltaic and photodetection characteristics, yielding a short-circuit current density of 39.61 mA $cm^{-2}$, an open-circuit voltage of 0.95 V, a fill factor of 86.74%, a power conversion efficiency of 32.68%, a responsivity of 0.72 A $W^{-1}$, and a detectivity of $1.78 \times 10^{16}$ Jones. These improvements are primarily attributed to enhanced carrier collection and strengthened built-in potential. Overall, this work provides an interpretable and accelerated framework for the design and optimization of next-generation high-efficiency integrated SC–PD devices.

**Acknowledgments**

The writers extremely appreciate Dr. Marc Burgelman, University of Gent, Belgium, for giving SCAPS simulation software.

*Corresponding Author: E-mail: jak_apee@ru.ac.bd (Jaker Hossain).

**Funding:** This work did not receive any funding from any funders

**Data availability:** Data will be available on request.

**Declaration of generative AI and AI-assisted technologies**

The authors declare that no AI or AI-assisted tools were used in preparing this manuscript.